\documentclass[aps,pre,amsmath,showpacs,floatfix,preprint]{revtex4}

\usepackage{graphicx}
\usepackage{amssymb}
\begin{document}

\title{Structural and conformational dynamics of \\
supercooled polymer melts:\\
insights from first-principles theory and simulations}
\author{Song-Ho Chong$^{1}$, Martin Aichele$^{2,3}$,
Hendrik Meyer$^{3}$, 
Matthias Fuchs$^{4}$, and
J\"org Baschnagel$^{3}$}
\affiliation{$^{1}$Institute for Molecular Science,
Okazaki 444-8585, Japan \\
$^{2}$Institut f\"ur Physik,
Johannes Gutenberg-Universit\"at,
55099 Mainz, Germany \\
$^{3}$Institut Charles Sadron, 6 rue Boussingault,
67083 Strasbourg, France \\
$^{4}$Fachbereich Physik, Universit\"at Konstanz,
78457 Konstanz, Germany}
\date{\today}

\begin{abstract}
We report on quantitative comparisons between simulation results of a
bead-spring model and mode-coupling theory calculations for the structural
and conformational dynamics of a supercooled,
unentangled polymer melt.  We find
semiquantitative agreement between simulation and theory,
except for processes that occur on intermediate length scales
between the compressibility plateau and the amorphous halo
of the static structure factor.
Our results suggest that the onset of slow relaxation
in a glass-forming melt can be
described in terms of monomer-caging
supplemented by chain connectivity.
Furthermore, a unified atomistic description of glassy
arrest and of conformational fluctuations that
(asymptotically) follow the Rouse model,
emerges from our theory.
\end{abstract}


\maketitle

\section{Introduction}

Polymeric melts can often be cooled down easily to vitrify 
into disordered solids. 
It is an important challenge for first-principles
approaches to develop an understanding of this
technologically important process.
In polymer science a further important challenge is to
derive well-known models of chain transport and relaxation in
melts~\cite{Crete2001}. In a melt, excluded volume interactions and
chain connectivity cause subdiffusive segment motion,
(approximately) described by the `Rouse' and `reptation' models
which consider a single chain in an effective field~\cite{Doi86}.
The challenge consists in deriving these models from microscopic
interactions.

In this paper, we propose an atomistic interpretation of the
structural and conformational dynamics of a bead-spring model for an
unentangled polymer melt~\cite{Bennemann98,Aichele01-all} by
quantitatively comparing simulation and first-principles
calculations. On the one hand, this explains the onset of the
viscous slowing-down, ultimately leading to kinetic arrest into an
amorphous solid (the glass
transition)~\cite{Arbe02,Lyulin03,Paul04,Baschnagel05}. On the other
hand, we find Rouse-like motion for very large chain length $N$, and
also explain characteristic deviations from pure Rouse behavior for
finite $N$. Our theory does not describe 
entanglements~\cite{Schweizer89-all} because we
start from isotropic, correlated monomer collisions which give
rise to the ``cage effect'' in dense fluids, but vanish in the limit of
infinitely thin chains where only topological constraints
(entanglements) are present. 
We aim to describe fundamental
consequences of the local steric packing in dense melts of flexible
polymers and thus, in a first step, neglect chemical structure like
torsional degrees of freedom; for  simulation studies of the glass
transition using chemically realistic models see,
e.g., Refs.~\cite{Lyulin03,Paul04,Paul06}.

Our approach is based on an extension of the
mode-coupling theory (MCT) for the glass transition~\cite{Goetze91b}
to polymer systems.
MCT predicts structural arrest --
also referred to as the idealized liquid-glass transition --
driven by the mutual blocking of a particle and its neighbors
at a critical temperature $T_{\rm c}$ which is located above the
glass-transition temperature $T_{\rm g}$.
Although complete structural arrest at $T_{\rm c}$
is not observed in experiments and simulations,
extensive tests of the theory carried out so far above $T_{\rm c}$
suggest that MCT deals properly with some essential features of
the structural relaxation in
glass-forming liquids~\cite{Goetze92,Goetze99}.

Our extension of the MCT to polymer systems will be done
following the site formalism~\cite{Schweizer97}.
In this formalism, each polymer molecule is divided into
interaction sites, corresponding to monomers or segments,
and the dynamics as well as the static structure
of polymers are characterized by site-site correlation functions.
For chains consisting of $N$ monomers, the site-site
correlation functions comprise $O(N^{2})$ elements,
and handling them is a formidable task for long chains.
A key assumption of our atomistic theory~\cite{Chong02c}
is to replace the site-specific intermolecular
surrounding of a monomer by an averaged one (equivalent-site
approximation~\cite{Schweizer97,Chong02c}),
while keeping the full intramolecular site-dependence.
For the statics this approximation has been verified by simulation
for wave lengths around the average segment separation
for a bead-spring model of a polymer melt~\cite{Aichele04}.
Here we extend the test by presenting
a quantitative comparison of the theory
for collective and single-chain dynamics with molecular-dynamics (MD)
simulations.
This test requires as only input the average static
structure factor $S(q)$, the average intrachain structure factor
$w(q)$, and the site-resolved intrachain structure factors
$w_{ab}(q)$ to be defined below.
These quantities are directly determined from the
simulation~\cite{Aichele04}.
Such a fully quantitative comparison
concerning the polymer dynamics is novel and, even for
systems of simper constituents, 
has been done only
recently~\cite{comparison-simpler-systems}.
It is motivated by successful MCT fits of the simulation data
for our model~\cite{Aichele01-all}.

It might be appropriate at the start to summarize the philosophy of
our approach for developing an atomistic theory of  the slow
structural dynamics in polymer melts, combining simulation and MCT.
We aim to derive first-principles results for measurable quantities
like intermediate scattering functions and mean-square
displacements. Within MCT, such averaged information on the
structural relaxation can be obtained from equilibrium structural
input up to a single unknown parameter, 
the time scale parameter, which needs to
be found from matching theory and simulation at one time instant
(we will choose the final relaxation
time of density fluctuations at the wave length corresponding to the
average segment separation for this purpose). 
We argue that quantities
obtained by averaging, assuming homogeneity,  should be studied
first to characterize macromolecular motion, postponing
consideration of heterogeneities and molecular/segmental variations
to future more detailed studies. A central question of our
investigation -- besides the one whether such an approach is
feasible at all -- is which static information is
required or sufficient to explain key features of macromolecular
motion. Therefore, we consider a simple polymer model without chemical
detail for which the required static
input can be obtained with high precision and
the theoretical predictions for the dynamics,
calculated without adjustable parameters, can be compared with
simulation results~\cite{comparison-simpler-systems}.
We expect to uncover fundamental mechanisms 
also present in real polymer melts, for which,
however, more complex static information than is necessary here
might be required as input to a theory for the dynamics.

The paper is organized as follows.
Section~\ref{sec:model} introduces the model polymer system
to be considered in the present article.
In Sec.~\ref{sec:static},
we briefly review static properties of the model.
Implications from the analysis of the static properties
will be employed in Sec.~\ref{sec:theory} to develop
a tractable theory for polymer dynamics.
In Sec.~\ref{sec:results}, theoretical predictions on
structural and conformational dynamics of the model
are compared with simulation results.
The paper is summarized in Sec.~\ref{sec:summary} with
some concluding remarks.
Appendix~\ref{appendix:MCT} is devoted to the derivation
of basic equations of motion, and
Appendix~\ref{appendix:Rouse-Fuchs}
to the derivation of the Rouse model
based on our microscopic approach.

\section{Model}
\label{sec:model}

We study a bead-spring model of linear chains, each containing
$N=10$ monomers of mass 
$m$~\cite{Bennemann98,Aichele01-all,Binder03,Baschnagel05}.
This is a model for highly flexible polymers, and
is among the simplest models exhibiting glassy arrest
and polymer specific dynamic anomalies. All monomers interact via a
truncated and shifted Lennard-Jones (LJ) potential
\begin{equation}
U_{\rm LJ}(r) =
\left\{
\begin{array}{lll}
\displaystyle{
4 \epsilon_{\rm LJ} \, [ \,
(\sigma_{\rm LJ}/r)^{12} - (\sigma_{\rm LJ}/r)^{6} \, ] + C} & \quad &
r < 2 \, r_{\rm min} \, , \\
0 & \quad &
r \ge 2 \, r_{\rm min} \, .
\end{array}
\right.
\end{equation}
In the following, all the quantities are expressed in
LJ units with the unit of length $\sigma_{\rm LJ}$,
the unit of energy $\epsilon_{\rm LJ}$ (setting Boltzmann's
constant $k_{\rm B}=1$), and the unit of time
$(m \sigma_{\rm LJ}^{2}/ \epsilon_{\rm LJ})^{1/2}$.
The constant $C = 127/4096$ is chosen so that $U_{\rm LJ}(r)$
vanishes continuously at $r = 2 \, r_{\rm min}$ with
$r_{\rm min} = 2^{1/6}$ being the minimum position of the
nontruncated potential.
In addition, successive monomers in a chain interact via a
finitely extendible nonlinear elastic (FENE)
potential~\cite{Kremer90}
\begin{equation}
U_{\rm FENE}(r) = - \frac{k}{2} \, R_{0}^{2} \ln
\left[ 1 - \biggr( \, \frac{r}{R_{0}} \, \biggr) \right],
\end{equation}
with $R_{0} = 1.5$ and $k = 30$.
The superposition of the LJ and FENE potentials leads to a
steep effective bond potential with a sharp minimum
at $r_{\rm b} = 0.9606$.

For this model we carried out MD simulations of polymer melts 
at constant pressure $p$ and constant temperature $T$. (The polymer melts comprise
100--120 chains, depending on temperature.)
The MD simulations were performed in two steps \cite{Bennemann98,Baschnagel05}:
For each $T$, the volume of the simulation box was first determined
in an isobaric simulation at $p=1$.
Then, this volume is kept fixed and the simulations
are continued in the canonical ensemble using
the Nos\'e-Hoover thermostat.
(The choice of this thermostat does not influence the relaxation dynamics of
the melt \cite{Bennemann98,Baschnagel05}.)
The simulations were carried out for the temperature
range $0.47 \le T \le 1$, corresponding to monomer densities
$0.91 \leq \rho_\mathrm{m} \leq 1.04$. The lowest simulated
temperature is slightly above 
$T_{\rm c}^{\rm MD} \approx 0.45$, the MCT critical
temperature as determined from the MD simulation
({\em cf}. Sec.~\ref{sec:theory-4}).
Before all measurements each state point $(T,\rho_\mathrm{m})$ is fully
equilibrated (the chains are allowed to diffuse several times over the distance
corresponding to their radius of gyration). For each state point, all quantities
are averaged over 150--200 independent time origins.
A more detailed description of the simulation technique
and simulation results for the model can be found in
Refs.~\cite{Bennemann98,Aichele01-all,Binder03,Baschnagel05}.

\section{Summary of static properties}
\label{sec:static}

Static structural and conformational properties of our model 
have been analyzed in detail in Ref.~\cite{Aichele04}. 
In this section, we briefly review
some of the main results of Ref.~\cite{Aichele04}, which help us to
develop a tractable theory for polymer dynamics.

\subsection{Static structure factors}
\label{sec:static-1}

Let us consider a polymer melt of $n$ chains, consisting of $N$
identical monomers, in a volume $V$. We denote by $\rho = n / V$ the
chain density. The static collective density fluctuations at
the monomer level can most naturally be characterized by the
monomer-monomer (or site-site) static structure factors
\begin{equation}
S_{ab}(q) = \frac{1}{n}
\langle
\rho_{a}({\bf q})^{*}
\rho_{b}({\bf q})
\rangle,
\label{eq:S-def}
\end{equation}
defined in terms of the coherent monomer-density fluctuations
for wave vector ${\bf q}$
\begin{equation}
\rho_{a}({\bf q}) = \sum_{i=1}^{n}
\exp [ i {\bf q} \cdot {\bf r}_{i}^{a} ] \quad
(a=1, \cdots, N).
\label{eq:rho-def}
\end{equation}
Here $\langle \cdot \rangle$ denotes the canonical averaging
for temperature $T$, and
${\bf r}_{i}^{a}$ represents the position of the $a$th monomer
in the $i$th chain.
Since the melt is spatially homogeneous and isotropic,
the structure factors depend only on the modulus of the
wave vector, $q = | {\bf q} |$.
One can split $S_{ab}(q)$ into intrachain and interchain parts
\begin{equation}
S_{ab}(q) = w_{ab}(q) + \rho h_{ab}(q),
\label{eq:Sab-decom}
\end{equation}
in which the intrachain contribution is given by
\begin{equation}
w_{ab}(q) = \frac{1}{n}
\Biggl\langle
\sum_{i=1}^{n} \exp [ - i {\bf q} \cdot
( {\bf r}_{i}^{a} - {\bf r}_{i}^{b} ) ]
\Biggr\rangle,
\end{equation}
and the interchain contribution by
\begin{equation}
\rho h_{ab}(q) = \frac{1}{n}
\Biggl\langle
\sum_{i \ne j}^{n} \exp [ - i {\bf q} \cdot
( {\bf r}_{i}^{a} - {\bf r}_{j}^{b} ) ]
\Biggr\rangle.
\end{equation}
These contributions reveal static correlations between monomers
belonging to the same chain or to different chains,
respectively.

Commonly, not these site-resolved quantities, but structure
factors averaged over all monomer pairs $(a,b)$
are discussed.
For instance,
we obtain the collective structure factor of the melt by
\begin{equation}
S(q) \equiv \frac{1}{N} \sum_{a,b=1}^{N} S_{ab}(q) =
\frac{1}{nN}
\langle
\rho_{\rm tot}({\bf q})^{*}
\rho_{\rm tot}({\bf q})
\rangle,
\label{eq:S-tot-def}
\end{equation}
which can be represented
in terms of the total monomer density fluctuations
(the second equality in the above equation)
\begin{equation}
\rho_{\rm tot}({\bf q}) \equiv \sum_{a=1}^{N} \rho_{a}({\bf q}) =
\sum_{i=1}^{n} \sum_{a=1}^{N}
\exp [ i {\bf q} \cdot {\bf r}_{i}^{a} ].
\label{eq:rho-tot-def}
\end{equation}
The average $S(q)$ can also be decomposed
into intrachain and interchain parts
\begin{equation}
S(q) = w(q) + \rho_{\rm m} h(q),
\label{eq:S-tot-decom}
\end{equation}
where $\rho_{\rm m} = N \rho$ denotes the monomer density, and
\begin{equation}
w(q) = \frac{1}{N} \sum_{a,b=1}^{N} w_{ab}(q), \quad
h(q) = \frac{1}{N^{2}} \sum_{a,b=1}^{N} h_{ab}(q).
\label{eq:w-h-tot-def}
\end{equation}
The intrachain contribution $w(q)$ is often called `form factor' in
the polymer literature~\cite{Doi86,Wittmer07}.

\begin{figure}[htb]
\includegraphics[width=0.5\linewidth]{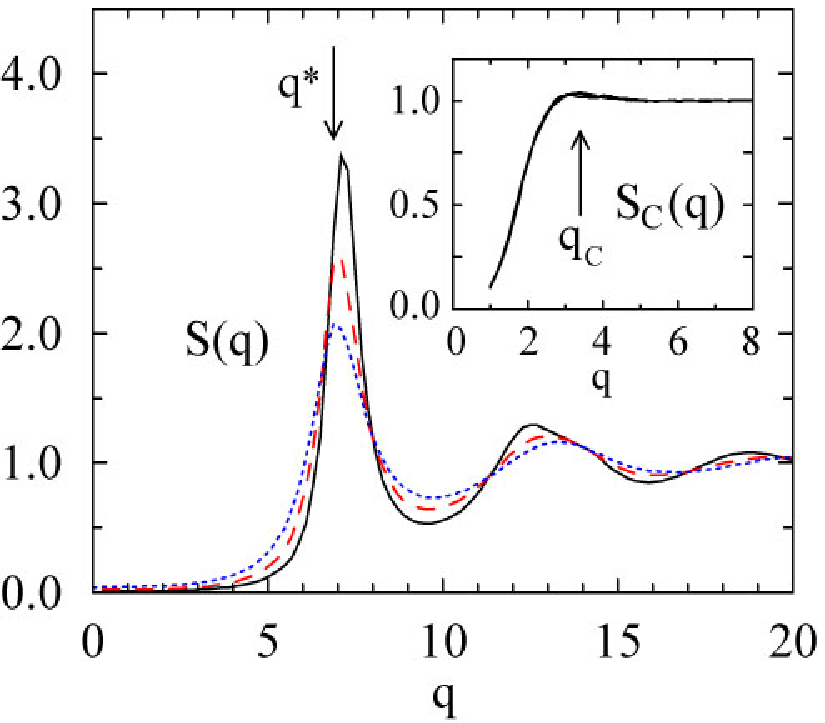}
\caption{
(Color online)
Collective static structure factor $S(q)$ of the melt as a function
of the modulus of the wave vector $q$ for temperatures
$T = 0.47$ (solid line), 0.70 (dashed line), and 1 (dotted line).
$S(q)$ exhibits a maximum around $q^{*} = 6.9$ whose position
is indicated by an arrow.
The inset shows $S_{\rm C}(q)$, the static structure factor
of the chain's center of mass, for $T = 0.47$, 0.70, and 1.
There is practically no temperature dependence in $S_{\rm C}(q)$,
and three curves cannot be distinguished from each other.
$S_{\rm C}(q)$ exhibits a weak maximum at $q_{\rm C} = 3.4$
whose position is indicated by an arrow.}
\label{fig:Sq}
\end{figure}

Figure~\ref{fig:Sq} shows the simulation result for $S(q)$ of our
model at $T = 0.47$, 0.70, and 1, which are representative
temperatures in the investigated range $0.47 \le T \le 1.0$. In this
$T$ interval, the collective structure of the melt is typical of a
dense disordered system. Due to the weak compressibility of the
melt, $S(q)$ is small in the $q \to 0$ limit. As $q$ increases,
$S(q)$ increases toward a maximum which occurs around $q^{*} = 6.9$
in our model. 
The corresponding length scale $2\pi/q^*$ is associated with the
average segment separation which is of the order of the effective monomer
diameter ($=1$ within the reduced units). Thus, the dominant
contribution to $S(q^{*})$ comes from the amorphous packing in
neighbor shells  around a monomer. Upon lowering
$T$, Fig.~\ref{fig:Sq} indicates that the packing becomes tighter,
which is reflected by the increased height of the peak $S(q^{*})$
and by the shift of its position $q^{*}$ to larger values. Such
changes of $S(q)$ at $q \approx q^{*}$ reflect the interchain
correlations since the intrachain structure factor $w(q)$ was found
to be nearly $T$ independent~\cite{Aichele04}.

In addition to monomer density fluctuations,
the static spatial arrangement of the center of mass (CM)
of chains and its $T$ dependence might be of interest.
The CM-CM static structure factor is defined by
\begin{equation}
S_{\rm C}(q) = \frac{1}{n}
\Biggl\langle
\sum_{i,j=1}^{n} \exp [ - i {\bf q} \cdot
( {\bf R}_{i} - {\bf R}_{j} ) ]
\Biggr\rangle,
\end{equation}
where ${\bf R}_{i}$ denotes the CM position of the $i$th chain. The
inset of Fig.~\ref{fig:Sq} shows the simulation result for $S_{\rm
C}(q)$. It is seen that $S_{\rm C}(q)$ is fairly featureless:
outside the small-$q$ regime reflecting the low compressibility of
the melt, $S_{\rm C}(q)$ quickly approaches the ideal gas behavior,
$S_{\rm C}(q) = 1$. There is a tiny peak at $q_{\rm C} \approx 3.4$,
whose height is, in contrast to that of $S(q)$, practically $T$
independent. 

One understands from Fig.~\ref{fig:Sq} that
the most pronounced $T$ dependence in the static structure
occurs in $S(q)$ around $q^{*}$
which reflect interchain monomer correlations.
This implies that the slowing-down of the dynamics of our model
upon lowering $T$
cannot result from static CM-CM or intrachain correlations, but
should be driven by interchain correlations at the monomer level,
i.e., by the nearest neighbors that are not directly bonded to each
other.
As we will see, this is one of the principal predictions
of our theory, according to which
the coherent dynamics close to $q^*$ enslaves all other dynamics,
including the CM and single-chain conformational dynamics.

\subsection{Equivalent-site approximation}
\label{sec:static-2}

In Ref.~\cite{Aichele04},  particular attention has been paid to the
dependence of static correlation functions on the position of the
monomer along the chain backbone to understand to what extent
specific monomer-monomer correlations deviate from the average
behavior. Since the site-site static correlation functions are
necessary input quantities for the mode-coupling approach based on
the site formalism, the comparison of the monomer-monomer
correlation functions with their monomer-averaged counterparts can
suggest suitable approximations and thus help developing a tractable
theory. In this and the next subsections, we summarize such
approximations which will be employed in Sec.~\ref{sec:theory}.

Let us introduce the site-site direct correlation
function $c_{ab}(q)$
via the site-site Ornstein-Zernike equation~\cite{Hansen86}
\begin{equation}
h_{ab}(q) = \sum_{x,y=1}^{N}
w_{ax}(q) \, c_{xy}(q) \,
[w_{yb}(q) + \rho h_{yb}(q)].
\label{eq:RISM}
\end{equation}
This is a generalized Ornstein-Zernike equation in which
intrachain correlations are accounted for through $w_{ab}(q)$;
it also serves as the defining equation of the direct correlation
function in terms of $w_{ab}(q)$ and $S_{ab}(q)$,
\begin{equation}
\rho c_{ab}(q) = w_{ab}^{-1}(q) - S_{ab}^{-1}(q).
\label{eq:c-def}
\end{equation}
Here $X_{ab}^{-1}(q)$
($X=w$ or $S$) denotes the $(a,b)$ element of the inverse of the
matrix ${\bf X}(q)$.

The difficulty in dealing with the site-site
correlation functions arises from the dependence on the
indices $(a,b)$.
Such functions consisting of $O(N^{2})$ elements cannot easily be
handled for large $N$.
One can argue, however, that, for long polymers,
chain end effects for interchain correlation functions
should be small,
suggesting to treat all sites of a homopolymer equivalently.
(This simplification is exact for a ring homopolymer.)
This equivalent-site approximation is usually
invoked for $c_{ab}(q)$, i.e.,
\begin{equation}
c(q) = c_{ab}(q) \quad (\mbox{equivalent-site approximation}).
\label{eq:PRISM-c}
\end{equation}
Equation~(\ref{eq:PRISM-c}) represents the principal idea of the
PRISM (polymer reference interaction site model) theory developed by
Schweizer, Curro, and coworkers~\cite{Schweizer97}.

Substituting Eq.~(\ref{eq:PRISM-c}) into
Eq.~(\ref{eq:RISM}) and then taking the summation $\sum_{a,b}$
of the resulting equation, one gets
the following scalar equation, called the PRISM equation,
in terms of the averaged quantities defined in
Eq.~(\ref{eq:w-h-tot-def}):
\begin{equation}
h(q) = w(q) \, c(q) \, [w(q) + \rho_{\rm m} h(q)].
\label{eq:PRISM}
\end{equation}
Equation~(\ref{eq:PRISM}) provides the following expression for $c(q)$
\begin{equation}
\rho_{\rm m} c(q) = 1/w(q) - 1/S(q),
\label{eq:PRISM-S1}
\end{equation}
in terms of the average $w(q)$ and $S(q)$.

The validity of the equivalent-site
approximation (\ref{eq:PRISM-c}) has been examined
for our model by comparing $c_{ab}(q)$ obtained from Eq.~(\ref{eq:c-def})
with $c(q)$ from Eq.~(\ref{eq:PRISM-S1}),
with the quantities on the right-hand sides of these
equations directly determined
from simulations (see Fig.~5 of Ref.~\cite{Aichele04}).
It has been demonstrated that the approximation is well
satisfied, except for functions involving the
chain ends.
This result suggests that, for our model, a theory
for the melt dynamics can be derived
by assuming Eq.~(\ref{eq:PRISM-c}) without introducing a
large error~\cite{comment-equivalent-site-app}.

\subsection{Additional ring approximation}
\label{sec:static-3}

Besides $c_{ab}(q)$, the static structure factors $S_{ab}(q)$
are necessary input quantities for the MCT based on the site
formalism ({\em cf.} Sec.~\ref{sec:theory}).
Thus, the equivalent-site
approximation~(\ref{eq:PRISM-c}) alone is insufficient
to obtain a tractable theory
since the specific monomer-position dependence still remains
in $S_{ab}(q)$.
This is obvious in view of the following relation
\begin{equation}
S_{ab}(q) = [ \{ {\bf I} - \rho \, {\bf w}(q) {\bf c}(q) \}^{-1}
{\bf w}(q) ]_{ab},
\end{equation}
which can be derived from Eqs.~(\ref{eq:Sab-decom}) and
(\ref{eq:RISM}).
Here ${\bf I}$ denotes the unit matrix.
Thus, even with the assumption $c_{ab}(q) = c(q)$,
a site dependence of $S_{ab}(q)$ results from chain connectivity,
i.e., from the matrix structure of $w_{ab}(q)$.
Therefore, it is desirable to have an additional approximation
which simplifies the treatment of $S_{ab}(q)$.

Remembering that the equivalent-site approximation~(\ref{eq:PRISM-c})
is exact for a ring polymer, we will derive an additional
approximation for linear chains based on another exact relation
for rings.
We will then examine the validity of this approximation
in our simulation.

A prominent feature of the site-site structure factor for a ring
polymer is that $\tilde{S}_{a}(q) \equiv \sum_{b=1}^{N} S_{ab}(q)$
is independent of $a$, and there holds
$\tilde{S}_{a}(q) = (1/N) \sum_{a=1}^{N}
\tilde{S}_{a}(q) = S(q)$; 
the second equality follows from
Eq.~(\ref{eq:S-tot-def}). Furthermore, from the identity $\sum_{x,b}
S_{ax}^{-1}(q) S_{xb}(q) = 1$, we also have $\tilde{S}^{-1}_{a}(q)
\equiv \sum_{b=1}^{N} S^{-1}_{ab}(q) = 1/S(q)$. In view of these
exact relations for a ring polymer, let us introduce the following
approximations for linear chains:
\begin{eqnarray}
& &
\tilde{S}_{a}(q) \equiv \sum_{b=1}^{N} S_{ab}(q) \approx S(q),
\label{eq:ring-app}
\\
& &
\tilde{S}^{-1}_{a}(q) \equiv \sum_{b=1}^{N} S^{-1}_{ab}(q)
\approx \frac{1}{S(q)}.
\label{eq:ring-app-2}
\end{eqnarray}

\begin{figure}[htb]
\includegraphics[width=0.5\linewidth]{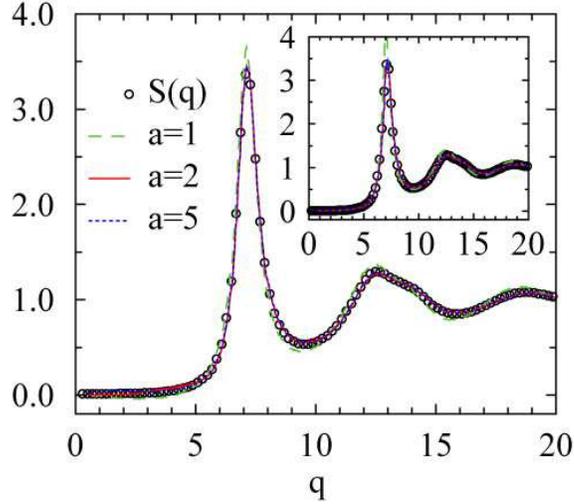}
\caption{
(Color online)
Comparison of the static structure factor $S(q)$ (circles)
with the site-dependent static structure factors $\tilde{S}_{a}(q)$
for $a=1$ (dashed line), $a=2$ (solid line), and $a=5$ (dotted line).
The inset compares $S(q)$ (circles) with
$1/\tilde{S}^{-1}_{a}(q)$
for $a=1$ (dashed line), $a=2$ (solid line), and $a=5$ (dotted line).
(The dotted lines for $a=5$ in the main panel and in the inset are
not clearly visible since they almost agree with the solid
lines for $a=2$.)
$\tilde{S}_{a}(q)$ and $\tilde{S}^{-1}_{a}(q)$
are defined by the first equality of
Eqs.~(\ref{eq:ring-app}) and (\ref{eq:ring-app-2}), respectively.
$S(q)$, $\tilde{S}_{a}(q)$, and $\tilde{S}^{-1}_{a}(q)$ are
taken from the simulation at $T = 0.47$.}
\label{fig:tilde-S}
\end{figure}

Figure~\ref{fig:tilde-S} and its inset examine to what extent the
ring approximations (\ref{eq:ring-app}) and (\ref{eq:ring-app-2})
hold for our model. It is seen that, except for $\tilde{S}_{1}(q)$
and $\tilde{S}^{-1}_{1}(q)$ referring to the end monomer, the ring
approximation is well satisfied, suggesting that this additional
approximation can also be used in deriving a theory for polymer
dynamics without introducing a large error. 
Let us add that both the equivalent-site and ring approximations
have been found to hold well
also for the liquid structure of some semiflexible
polymer models~\cite{Vettorel07}.

\section{Theory}
\label{sec:theory}

\subsection{MCT equations for coherent structural dynamics}
\label{sec:theory-1}

In the site formalism,
collective structural dynamics are to be described by
site-site density correlators
\begin{equation}
F_{ab}(q,t) = \frac{1}{n}
\langle \rho_{a}({\bf q})^{*}
e^{i {\cal L} t}
\rho_{b}({\bf q}) \rangle \quad
(a,b = 1, \cdots, N),
\label{eq:Fab-def}
\end{equation}
whose initial values are the static structure factors
$S_{ab}(q) = F_{ab}(q,0)$.
Here, ${\cal L}$ denotes the Liouville operator
appropriate for Newtonian dynamics. 
MCT equations of motion for $F_{ab}(q,t)$ for general flexible molecules
are derived in Appendix~\ref{appendix:MCT}, and
consist of the Zwanzig-Mori exact equation of motion
and an approximate expression for the memory kernel.
The former is obtained by introducing a
projection operator ${\cal P}$ onto the monomer-density
fluctuations and the corresponding longitudinal current 
fluctuations, and reads
[{\em cf.} Eq.~(\ref{eq:GLE-a})]:
\begin{equation}
\partial_{t}^{2} F_{ab}(q,t) +
\sum_{x=1}^{N} \Omega_{ax}^{2}(q) \, F_{xb}(q,t) +
\sum_{x=1}^{N} \int_{0}^{t} dt' \, M_{ax}(q,t-t') \,
\partial_{t'} F_{xb}(q,t') = 0.
\label{eq:GLE}
\end{equation}
Here $\Omega_{ab}^{2}(q)$ represents the characteristic
frequency given by
\begin{equation}
\Omega_{ab}^{2}(q) = q^{2} v^{2} S^{-1}_{ab}(q),
\label{eq:Omega}
\end{equation}
with $v^{2} = k_{\rm B}T / m$ ($=T$ within the reduced units)
denoting the monomer thermal velocity, and a formally 
exact expression for the memory kernel reads
\begin{equation}
M_{ab}(q,t) =
\frac{1}{nv^{2}}
\langle f_{a}({\bf q})^{*}
\exp (i {\cal QLQ} t)
f_{b}({\bf q}) \rangle,
\end{equation}
in terms of the fluctuating random force
$f_{a}({\bf q})$ 
which evolves with the generator ${\cal QLQ}$, where ${\cal Q}
\equiv 1 - {\cal P}$. 
An approximate expression for $f_{a}({\bf q})$
as derived under the mode-coupling approach 
in Appendix~\ref{appendix:MCT} is given by
[omitting the irrelevant factor $-i$ from Eq.~(\ref{eq:MCT-random-force})]
\begin{equation}
f_{a}({\bf q}) = \frac{\rho v^{2}}{n}
\sum_{\bf k} \sum_{x=1}^{N}
(\widehat{\bf q} \cdot {\bf k}) \,
c_{xa}(k) \, \rho_{x}({\bf k}) \, \rho_{a}({\bf p}),
\label{eq:f-orig}
\end{equation}
in which $\widehat{\bf q} = {\bf q} / q$ and ${\bf p} = {\bf q} - {\bf k}$.
With the use of the factorization approximation~(\ref{eq:factori-appendix}),
one finally arrives at the following MCT expression
for the kernel
[{\em cf.} Eq.~(\ref{eq:appendix-MCT})]:
\begin{eqnarray}
M_{ab}(q,t) &=&
\frac{\rho v^{2}}{(2 \pi)^{3}}
\sum_{x,y=1}^{N}
\int d{\bf k} \,
\bigl\{ \,
(\widehat{\bf q} \cdot {\bf k})^{2} c_{ax}(k) c_{by}(k)
F_{xy}(k,t) F_{ab}(p,t) +
\nonumber \\
& & \qquad \qquad \qquad \qquad \quad
+ \,
(\widehat{\bf q} \cdot {\bf k}) (\widehat{\bf q} \cdot {\bf p})
c_{ax}(k) c_{by}(p)
F_{xb}(k,t) F_{ay}(p,t)
\, \bigr\} \, .
\label{eq:MCT}
\end{eqnarray}

Equations (\ref{eq:GLE}) and (\ref{eq:MCT})
provide a set of closed equations for determining
site-site coherent density correlators
$F_{ab}(q,t)$, provided static quantities
$S_{ab}(q)$ and $c_{ab}(q)$ are known.
From a computational point of view, however, it is quite
demanding to solve these $N \times N$ matrix
equations since $N$ may become large for polymeric 
systems~\cite{comment-computational-time}.
It is at this point where the analysis of the static properties,
presented in Ref.~\cite{Aichele04} and summarized in Sec.~\ref{sec:static},
will help us to develop further approximations.

As mentioned in Sec.~\ref{sec:static-2},
the equivalent-site approximation,
$c_{ab}(q) = c(q)$, is well justified for our model.
So, we insert $c_{ab}(q) = c(q)$ in Eq.~(\ref{eq:f-orig}) and obtain
\begin{equation}
f_{a}({\bf q}) = \frac{\rho v^{2}}{n}
\sum_{\bf k} (\widehat{\bf q} \cdot {\bf k}) \, c(k) \,
\rho_{\rm tot}({\bf k}) \, \rho_{a}({\bf p}),
\label{eq:f-intermediate}
\end{equation}
where $\rho_{\rm tot}({\bf k})$ denotes the total monomer density
fluctuations introduced in Eq.~(\ref{eq:rho-tot-def}). This
expression reveals that the equivalent-site approximation alone does
not suffice to simplify the problem: the dependence of $f_{a}({\bf
q})$ on the monomer position remains, and the resulting
$M_{ab}(q,t)$ still carries the $(a,b)$ dependence, i.e., it
consists of $O(N^{2})$ elements. Furthermore, no simplification is
yet achieved concerning the frequency matrix (\ref{eq:Omega}).

Progress is made if we invoke the second approximation
described in Sec.~\ref{sec:static-3}.
The frequency term (\ref{eq:Omega}) can be simplified
by the use of the ring approximation (\ref{eq:ring-app-2}) to
\begin{equation}
\sum_{a=1}^{N} \Omega^{2}_{ab}(q) =
q^{2} v^{2} \sum_{a=1}^{N} S^{-1}_{ab}(q) \approx
q^{2} v^{2} / S(q) \equiv \Omega^{2}(q).
\label{eq:Omega-ring}
\end{equation}
A corresponding simplification can be introduced
for the fluctuating force $f_{a}({\bf q})$.
To this end,
we notice that $f_{a}({\bf q})$ 
in Eq.~(\ref{eq:f-intermediate}) originates from
interchain interactions represented through the
direct correlation function.
It is then reasonable to expect that, for long chains, the specific
monomer-position dependence in $f_{a}({\bf q})$ is small,
and can be well approximated by an averaged one.
(This approximation is exact for rings.)
We therefore introduce an approximation
\begin{equation}
f_{a}({\bf q}) \approx \frac{1}{N} \sum_{a=1}^{N}
f_{a}({\bf q}).
\end{equation}
This ring approximation replaces the site-specific surroundings
of a monomer by an averaged one, and then the fluctuating force
is given by
\begin{equation}
f_{a}({\bf q}) =
\frac{\rho v^{2}}{n N}
\sum_{\bf k}
(\widehat{\bf q} \cdot {\bf k}) \, c(k) \,
\rho_{\rm tot}({\bf k}) \,
\rho_{\rm tot}({\bf p}).
\label{eq:f-final}
\end{equation}
This leads to an expression for $M_{ab}(q,t)$ (to be summarized below)
which now does not depend on the site indices $(a,b)$.

The approximations discussed so far allow us to derive a
set of closed MCT
equations for the collective total monomer density correlators
\begin{equation}
F(q,t) \equiv \frac{1}{N} \sum_{a,b=1}^{N} F_{ab}(q,t) =
\frac{1}{nN}
\langle
\rho_{\rm tot}({\bf q})^{*} e^{i {\cal L} t}
\rho_{\rm tot}({\bf q})
\rangle,
\label{eq:F-def-2}
\end{equation}
whose initial value is $F(q,0) = S(q)$
[{\em cf.} Eq.~(\ref{eq:S-tot-def})].
To this end, we take $(1/N) \sum_{a,b}$ of Eq.~(\ref{eq:GLE}),
and then insert the frequency term (\ref{eq:Omega-ring})
and the memory kernel with the
fluctuating force given in Eq.~(\ref{eq:f-final})
under the factorization approximation~(\ref{eq:factori-appendix}).
This gives the following set of MCT equations for the
normalized coherent density correlators
$\phi(q,t) \equiv F(q,t) / S(q)$:
\begin{eqnarray}
& &
\partial_{t}^{2} \phi(q,t) + \Omega^{2}(q) \, \phi(q,t) +
\Omega^{2}(q) \int_{0}^{t} dt^{\prime} \,
m(q,t-t^{\prime}) \, \partial_{t^{\prime}} \phi(q,t^{\prime}) = 0,
\label{eq:GLE-final} \\
& &
m(q,t) =
\frac{1}{2} \int d{\bf k} \, V({\bf q}; {\bf k}, {\bf p}) \,
\phi(k,t) \, \phi(p,t).
\label{eq:MCT-final}
\end{eqnarray}
Here $\Omega^{2}(q) = q^{2} v^{2} / S(q)$, and the vertex function
reads
\begin{equation}
V({\bf q}; {\bf k}, {\bf p}) =
\frac{\rho_{\rm m}}{(2 \pi)^{3} q^{2}} \,
S(q) S(k) S(p)
\left\{ \widehat{\bf q} \cdot [{\bf k} c(k) + {\bf p} c(p)] \right\}^{2}.
\label{eq:vertex}
\end{equation}
One can solve these equations for $\phi(q,t)$ provided
the average static quantities $S(q)$ and $c(q)$ are given as input.

Equations~(\ref{eq:GLE-final}) and (\ref{eq:MCT-final})
merit some comments:
(i) These equations are formally identical to MCT equations
for monatomic liquids.
Polymer-specific effects, such as local stiffness of the
chain backbone or chain length $N$, enter the relaxation
only via the direct correlation function $c(q)$,
the structure factor $S(q)$,
and the monomer density $\rho_{\rm m}$.
These static equilibrium features fully determine the long-time
coherent dynamics of the melt.
(ii) Equations~(\ref{eq:MCT-final}) and (\ref{eq:vertex})
indicate that the memory kernel
contains the factor $S(q) S(k) S(p)$.
So, the slow dynamics upon lowering $T$
should be mainly driven by wave vectors close to $q^{*}$
because there, $S(q)$ is largest and the strongest
dependence on $T$ occurs ({\em cf.} Sec.~\ref{sec:static-2}).
Thus, our theory predicts that the glassy structural
slowing-down is connected to the increase of the first peak of $S(q)$,
i.e., to the local cage effect.

We finally notice that the so-called regular contribution
to the memory kernel~\cite{Goetze91b} is discarded in our theory,
and our approximate memory kernel is completely
given by the mode-coupling expression.
The latter provides the slow contribution relevant
for the structural slowing-down.
The regular contribution is supposed to embody memory effects
already present in the normal high-$T$ state of liquids,
and accounts for the fast dynamics in the short-time regime.
We drop the regular contribution since
it does not affect the MCT predictions on the
slow-relaxation regime~\cite{Goetze91b,Kob02}.
Thus, care has to be taken in comparing theoretical
predictions with simulation results,
since the theory without the regular contribution
does not properly describe the short-time
dynamics~\cite{comment-regular-part}.

\subsection{MCT equations for single-chain dynamics}
\label{sec:theory-2}

The basic variable characterizing the dynamics 
of a single (or tagged) chain is
\begin{equation}
\rho_{a}^{\mathrm s}({\bf q},t) =
e^{ i {\bf q} \cdot {\bf r}_{\mathrm s}^{a}(t)},
\label{eq:rho-s-def}
\end{equation}
where ${\bf r}_{\mathrm s}^{a}(t)$ denotes the position of $a$th
monomer in the tagged (labeled ${\rm s}$) chain
at time $t$.
The density correlator for the single-chain dynamics is defined by
\begin{equation}
F_{ab}^{\mathrm s}(q,t) =
\langle \rho_{a}^{\mathrm s}({\bf q})^{*}
\rho_{b}^{\mathrm s}({\bf q},t)
\rangle,
\label{eq:Fs-def}
\end{equation}
whose initial value is the intrachain structure factor,
$w_{ab}(q) = F_{ab}^{\mathrm s}(q,0)$.

The derivation of
the MCT equations for $F_{ab}^{\mathrm s}(q,t)$
is outlined in Appendix~\ref{appendix:MCT-3},
and the resulting matrix equations can be summarized as
\begin{equation}
\partial_{t}^{2} F_{ab}^{\mathrm s}(q,t) +
\sum_{x=1}^{N} \Omega_{ax}^{{\mathrm s} \, 2}(q) \,
F_{xb}^{\mathrm s}(q,t) +
\sum_{x,y=1}^{N} \Omega_{ax}^{{\mathrm s} \, 2}(q) \,
\int_{0}^{t} dt' \,
m_{xy}^{\mathrm s}(q,t-t') \,
\partial_{t'} F_{yb}^{\mathrm s}(q,t') = 0,
\label{eq:GLE-s}
\end{equation}
where the frequency matrix is given by
\begin{equation}
\Omega_{ab}^{{\mathrm s} \, 2}(q) =
q^{2} v^{2} w^{-1}_{ab}(q),
\label{eq:GLE-s-b}
\end{equation}
and the MCT expression for $m^{\mathrm s}_{ab}(q,t)$
under the equivalent-site approximation (\ref{eq:PRISM-c})
reads
\begin{equation}
m^{\mathrm s}_{ab}(q,t) =
\sum_{x=1}^{N} w_{ax}(q)
\int d{\bf k} \, V^{\mathrm s}({\bf q}; {\bf k}, {\bf p}) \,
F_{xb}^{\mathrm s}(k,t) \, \phi(p,t),
\label{eq:MCT-s}
\end{equation}
with the vertex function
\begin{equation}
V^{\mathrm s}({\bf q}; {\bf k}, {\bf p}) =
\frac{\rho_{\rm m}}{(2 \pi)^{3} q^{2}}
S(p) \, (\widehat{\bf q} \cdot {\bf p})^{2} \, c(p)^{2}.
\label{eq:vertex-s}
\end{equation}
Equations~(\ref{eq:GLE-s}) and (\ref{eq:MCT-s}) constitute a set of closed $N \times N$-matrix MCT equations for
the single-chain density correlators $F_{ab}^{\mathrm s}(q,t)$.
One can solve these equations with the knowledge of the
static quantities -- $S(q)$, $c(q)$, and $w_{ab}(q)$ --
and of the coherent density correlators $\phi(q,t)$.
It is clear from Eq.~(\ref{eq:MCT-s}) that the slowing-down
of the single-chain dynamics is driven by that of the
coherent dynamics.

Unlike the MCT equations for the coherent dynamics,
one cannot simplify the matrix structure of
Eqs.~(\ref{eq:GLE-s})--(\ref{eq:MCT-s}) for
the single-chain density correlators $F_{ab}^{\mathrm s}(q,t)$
in order to properly describe chain-connectivity effects,
taken into account through the
intrachain structure factor matrix $w_{ab}(q)$.
For example,
one needs the site-site $F_{ab}^{\mathrm s}(q,t)$ to fully describe
the chain conformational dynamics, i.e., all the Rouse-mode 
correlators introduced below,
and has to solve the matrix MCT equations
(\ref{eq:GLE-s}) and (\ref{eq:MCT-s}) for this purpose.
Let us note in this connection that,
from a computational point of view,
it is not so demanding to solve these matrix MCT equations
for $F_{ab}^{\mathrm s}(q,t)$.
This is because the most time consuming part
in numerically solving the MCT equations is spent
in solving the ones for 
coherent dynamics~\cite{comment-coherent-vs-incoherent}.

\subsection{MCT equations for Rouse-mode correlators}
\label{sec:theory-3}

In this subsection, the site-density description of the single-chain
dynamics shall be related to the traditional Rouse description, and
we derive the MCT equations for the Rouse-mode correlators. 
Let us stress that the present rewriting is exact and always possible.
The $N$ degrees of freedom of segmental motion are mapped onto $N$
modes labeled by $p$. If the Rouse model holds, the modes will be
statistically independent and the matrix of Rouse-mode correlators
introduced below becomes diagonal.

Let us introduce the following $N \times N$
orthogonal matrix $P_{ap}$ ($a = 1,2,\cdots,N$ and $p =
0,1,\cdots,N-1$):
\begin{equation}
P_{ap} =
\left\{
\begin{array}{ll}
\displaystyle{ \sqrt{\frac{1}{N}} } &
(p=0), \\
\displaystyle{ \sqrt{\frac{2}{N}} \,
\cos \left[ \frac{(a-1/2) p \pi}{N} \right] } &
(p = 1,2,\cdots,N-1). \\
\end{array}
\right.
\label{eq:P-def}
\end{equation}
The Rouse-mode vectors shall be defined
in terms of the monomer positions as
${\bf X}_{p}(t) = \sum_{a=1}^{N}
P_{ap} {\bf r}_{\mathrm s}^{a}(t)$~\cite{Verdier66}, i.e.,
\begin{equation}
{\bf X}_{p}(t) =
\left\{
\begin{array}{ll}
\displaystyle{ \sqrt{\frac{1}{N}}
\sum_{a=1}^{N} {\bf r}_{\mathrm s}^{a}(t) } \quad &
(p=0), \\
\displaystyle{ \sqrt{\frac{2}{N}}
\sum_{a=1}^{N} {\bf r}_{\mathrm s}^{a}(t) \,
\cos \left[ \frac{(a-1/2) p \pi}{N} \right] } \quad &
(p = 1,2,\cdots,N-1). \\
\end{array}
\right.
\label{eq:Rouse-1}
\end{equation}
The inverse relation is given by
${\bf r}_{\mathrm s}^{a}(t) = \sum_{p=0}^{N-1} P_{ap} {\bf X}_{p}(t)$:
\begin{equation}
{\bf r}_{\mathrm s}^{a}(t) = \sqrt{\frac{1}{N}}
\left\{
{\bf X}_{0}(t) +
\sqrt{2} \sum_{p=1}^{N-1} {\bf X}_{p}(t) \,
\cos \left[ \frac{(a-1/2) p \pi}{N} \right]
\right\}.
\label{eq:Rouse-2}
\end{equation}
We introduce the Rouse-mode correlators as
\begin{equation}
C_{p p'}(t) =
\left\{
\begin{array}{ll}
\displaystyle{
[\langle {\bf X}_{0}(0) \cdot {\bf X}_{0}(t) \rangle -
\langle {\bf X}_{0}(0) \cdot {\bf X}_{0}(0) \rangle] \, / \, 3N} \quad &
(p=p'=0), \\
\displaystyle{
[\langle {\bf X}_{0}(0) \cdot {\bf X}_{p'}(t) \rangle -
\langle {\bf X}_{0}(0) \cdot {\bf X}_{p'}(0) \rangle] \, / \, 3N} \quad &
(p=0, \, p' \ne 0), \\
\displaystyle{
\langle {\bf X}_{p}(0) \cdot {\bf X}_{p'}(t) \rangle \, / \, 3N}
\quad &
(p \ne 0, \, p' \ne 0). \\
\end{array}
\right.
\label{eq:Rouse-3}
\end{equation}
From the definition, it is obvious that
\begin{equation}
C_{00}(0) = 0, \quad
C_{0p}(0) = 0 \, \, \mbox{ for } \, \, p>0.
\label{eq:C0-1}
\end{equation}
Let us introduce the following $(N-1) \times (N-1)$ matrix
to denote the initial values of $C_{pp'}(t)$ for $p,p'>0$:
\begin{equation}
\widehat{C}_{pp'} \equiv C_{pp'}(0)
\,\, \mbox{ defined only for } \,\,
p,p'>0.
\label{eq:C0-2}
\end{equation}
It is necessary to introduce this new matrix in order to
discuss the inverse matrix of $C_{pp'}(0)$:
the inverse of the $N \times N$ matrix $C_{pp'}(0)$
does not exist because of Eq.~(\ref{eq:C0-1}),
whereas that of the $(N-1) \times (N-1)$ matrix
$\widehat{C}_{pp'}$ does exist, and shall be denoted
as $\widehat{C}_{pp'}^{-1}$.

Since ${\bf X}_{0}(t) = \sqrt{N} {\bf R}_{\mathrm s}(t)$
with ${\bf R}_{\mathrm s}(t)$ denoting the CM position
of the tagged chain at time $t$,
$C_{00}(t)$ is related to the CM
mean-square displacement (MSD)
$g_{\rm C}(t) =
\langle [ {\bf R}_{\mathrm s}(t) - {\bf R}_{\mathrm s}(0) ]^{2} \rangle$:
\begin{equation}
g_{\rm C}(t) = - 6 \, C_{00}(t).
\label{eq:MSD-COM}
\end{equation}
For the monomer MSD,
$g_{a}(t) =
\langle [{\bf r}_{\mathrm s}^{a}(t)-{\bf r}_{\mathrm s}^{a}(0)]^{2} \rangle$,
it follows from Eqs.~(\ref{eq:Rouse-2}) and (\ref{eq:Rouse-3}) that
\begin{eqnarray}
g_{a}(t) &=&
g_{\rm C}(t) -
12 \sqrt{2} \sum_{p'=1}^{N-1} C_{0p'}(t)
\cos \left[ \frac{(a-1/2) p' \pi}{N} \right]
\nonumber \\
& & \quad
+ \,
12 \sum_{p=1}^{N-1} \sum_{p'=1}^{N-1}
\left[ C_{pp'}(0) - C_{pp'}(t) \right] \,
\cos \left[\frac{(a-1/2) p \pi}{N} \right]
\cos \left[\frac{(a-1/2) p' \pi}{N} \right].
\label{eq:MSD-monomer}
\end{eqnarray}
In particular, one gets for the MSD averaged over all the monomers
in a chain
\begin{equation}
g_{\rm M}(t) \equiv (1/N) \sum_{a=1}^{N} g_{a}(t) =
g_{\rm C}(t) +
6 \sum_{p=1}^{N-1}
[\, C_{pp}(0) - C_{pp}(t) \,].
\label{eq:MSD-monomer-averaged}
\end{equation}

We next consider how the Rouse-mode correlators $C_{pp'}(t)$
are related to the site-density correlators $F_{ab}^{\mathrm s}(q,t)$.
This will allow us to write down MCT equations for the
former based on the ones for the latter.
Since the density fluctuations of the tagged chain
for small ${\bf q}$ are given by
$\rho_{a}^{\mathrm s}({\bf q},t) \approx
1 + i {\bf q} \cdot {\bf r}_{\mathrm s}^{a}(t)$,
it is easily understood that
$C_{pp'}(t)$ can be expressed as a linear
combination of $F_{ab}^{\mathrm s}(q,t)$ for $q \to 0$.
Indeed, one can show that
\begin{equation}
\frac{1}{N} \sum_{a,b=1}^{N}
P_{ap} F_{ab}^{\mathrm s}(q \to 0,t) P_{bp'} =
\left\{
\begin{array}{ll}
1 + q^{2} [ C_{00}(t) - A ] + O(q^{4})   \quad & (p=p'=0),    \\
q^{2} [ C_{0p'}(t) - B_{p'} ] + O(q^{4})       & (p=0, p'>0), \\
q^{2} C_{pp'}(t) + O(q^{4})                    & (p,p'>0),
\end{array}
\right.
\label{eq:Fqs-Rouse-small-q}
\end{equation}
where
$A = (1/6N^{2})
\sum_{a,b=1}^{N} \langle
({\bf r}_{\mathrm s}^{a} - {\bf r}_{\mathrm s}^{b})^{2}
\rangle = R_{\rm g}^{2}/3$
with $R_{\rm g}$ denoting the radius of gyration of a chain,
and
$B_{p} = (1/3\sqrt{2} N^{2})
\sum_{a,b=1}^{N} \langle
({\bf r}_{\mathrm s}^{a} - {\bf r}_{\mathrm s}^{b})^{2}
\rangle
\cos [(b-1/2) p \pi / N ]$.
Using these relations, one derives the following MCT equations
for $C_{pp'}(t)$ by taking the $q \to 0$ limit of
Eqs.~(\ref{eq:GLE-s})--(\ref{eq:vertex-s}):
\begin{equation}
\partial_{t}^{2} C_{pp'}(t) +
v^{2} D_{pp'} +
v^{2} \sum_{p''=0}^{N-1} E_{pp''} C_{p''p'}(t) +
v^{2}
\sum_{p''=0}^{N-1}
\int_{0}^{t} dt' \,
m_{pp''}(t-t') \, \partial_{t'} C_{p''p'}(t') = 0,
\label{eq:GLE-Rouse-1-a}
\end{equation}
where
\begin{eqnarray}
D_{pp'} &=& \delta_{p0} \delta_{p'0} / N,
\label{eq:GLE-Rouse-1-b}
\\
E_{pp'} &=&
\left\{
\begin{array}{ll}
\widehat{C}_{pp'}^{-1} / N \quad & (p,p'>0),           \\
0                                & (\mbox{otherwise}).
\end{array}
\right.
\label{eq:GLE-Rouse-1-c}
\end{eqnarray}
and the expression for the memory kernel reads
\begin{equation}
m_{pp'}(t) = \frac{\rho_{\rm m}}{6 \pi^{2}}
\int dk \, k^{4} S(k) c(k)^{2} \,
\left[
  \sum_{a,b=1}^{N} P_{ap} F_{ab}^{\mathrm s}(k,t) P_{bp'} 
\right] \, 
\phi(k,t).
\label{eq:MCT-Rouse}
\end{equation}
These MCT equations for $C_{pp'}(t)$
can be solved provided the static quantities --
$S(q)$, $c(q)$, $w_{ab}(q)$, and $\widehat{C}_{pp'}$ --
and the full site-site single-chain density correlators
$F_{ab}^{\mathrm s}(q,t)$
as well as the coherent density correlators $\phi(q,t)$ are known.
MSDs can then be obtained from
Eqs.~(\ref{eq:MSD-COM}) through (\ref{eq:MSD-monomer-averaged}).
Again, it is clear from Eq.~(\ref{eq:MCT-Rouse}) that
the slowing-down of the Rouse-mode dynamics and MSDs is
dictated by that of the coherent dynamics.

In Appendix~\ref{appendix:Rouse-Fuchs},
we will show how the Rouse model emerges from our MCT equations
in the asymptotic limit of large $N$.

\subsection{Universal MCT predictions}
\label{sec:theory-4}

Here, we briefly summarize some universal MCT predictions
which are necessary for understanding the present paper.
As described in Ref.~\cite{Chong02}, all universal results
concerning the MCT-liquid-glass transition dynamics,
originally developed for simple systems~\cite{Goetze91b},
are also valid for molecular systems,
and the MCT for polymer melts developed
in Sec.~\ref{sec:theory} shares this feature.
This justifies the use of the MCT universal predictions
in analyzing polymer data, whose validity has been
tacitly assumed in previous studies of our
model~\cite{BennemannBaschnagelPaul1999_incoherent,Aichele01-all,Baschnagel05}.

One of the central predictions of MCT is the existence
of a critical temperature $T_{\rm c}$.
The long-time limit, or the nonergodicity parameter,
of the coherent density correlator
$f(q) \equiv \phi(q,t \to \infty)$
obeys the implicit equation
\begin{equation}
\frac{f(q)}{1-f(q)} = {\cal F}_{q}[f],
\label{eq:NEP}
\end{equation}
which can be derived by taking the $t \to \infty$ limit of
Eqs.~(\ref{eq:GLE-final}) and (\ref{eq:MCT-final})
and introducing the mode-coupling functional
${\cal F}_{q}[f] \equiv (1/2) \int d {\bf k}
V({\bf q};{\bf k},{\bf p}) f(k) f(p)$.
One gets trivial solutions $f(q) = 0$ for $T > T_{\rm c}$,
meaning that the density fluctuations relax completely
at long times, a characteristic feature of the
ergodic liquid state.
On the other hand, nontrivial solutions $f(q) > 0$
can be obtained for $T \le T_{\rm c}$,
describing nonergodic dynamics in which
density fluctuations cannot fully decay.
The nonergodicity parameter $f(q)$ measures the ``solidity'' of
such an amorphous solid on length scales $\approx 2 \pi / q$,
and is thus also referred to as the glass-form factor
or the Debye-Waller factor.
The ergodic-to-nonergodic transition at $T_{\rm c}$ is
called the idealized glass transition, and
$f(q)$ at $T = T_{\rm c}$, to be denoted as $f^{\rm c}(q)$,
is referred to as the critical nonergodicity parameter.
It also has the meaning of the plateau height
in the two-step relaxation of
$\phi(q,t)$, and quantifies the strength of its $\alpha$
relaxation (see below).

MCT predicts that for temperatures close to but above $T_{\rm c}$,
to which we restrict our attention in the following,
the dynamics of any time-dependent correlation
function $\phi_{\rm X}(t)$ coupling to density fluctuations
exhibits a two-step relaxation:
the relaxation toward the plateau,
and the final relaxation from the plateau to zero.
These relaxations are respectively
characterized by the time scales $t_{\sigma}$ and $t_{\sigma}^{\prime}$
defined by
\begin{eqnarray}
t_{\sigma} &=& t_{0} \, / \, | \, \sigma \, |^{\delta},
\quad \delta = 1/2a,
\label{eq:t-sigma}
\\
t_{\sigma}^{\prime} &=& t_{0} \, B^{-1/b} \, / \, | \, \sigma \, |^{\gamma},
\quad \gamma = (1/2a) + (1/2b).
\label{eq:t-sigma-prime}
\end{eqnarray}
Here $t_{0}$ denotes some microscopic time scale~\cite{Franosch98}.
$\sigma$ is called the separation parameter, and measures
the distance from the critical point:
\begin{equation}
\sigma = C \, \epsilon, \quad
\epsilon = (T_{\rm c} - T) \, / \, T_{\rm c}.
\end{equation}
Except for $t_{0}$, all the exponents and 
the constants $B$ and $C$ appearing in
these equations can be evaluated
from the mode-coupling functional ${\cal F}_{q}[f]$
with the knowledge of $f^{\rm c}(q)$~\cite{Goetze91b,Franosch97-Fuchs98}.
According to MCT, the dynamics dramatically slows down upon
lowering $T$ since the time scales $t_{\sigma}$ and
$t_{\sigma}^{\prime}$ diverge for $T \to T_{\rm c}+$.
The two-step-relaxation scenario emerges
since the $T$ dependence of $t_{\sigma}^{\prime}$ is
stronger than that of $t_{\sigma}$.

The dynamics which occurs near the plateau is referred to
as the $\beta$ process.
The height of the plateau is given by the critical
nonergodicity parameter $f_{\rm X}^{\rm c}$ of
the correlator $\phi_{\rm X}(t)$, which can be determined from the
corresponding mode-coupling functional
via an equation analogous to Eq.~(\ref{eq:NEP}).
MCT predicts that there holds for $\sigma \to -0$
(i.e., for $T \to T_{\rm c}+$)
\begin{equation}
\phi_{\rm X}(t) = f_{\rm X}^{\rm c} + h_{\rm X}
\sqrt{|\sigma|} \, g(t/t_{\sigma}) + O(\sigma).
\label{eq:factorization-theorem}
\end{equation}
Here $h_{\rm X}$ is called the critical amplitude
and $g(\hat{t})$ the $\beta$ correlator~\cite{Goetze91b}.
Equation~(\ref{eq:factorization-theorem})
is called the factorization theorem,
according to which the dependence of the correlator $\phi_{\rm X}(t)$
on ${\rm X}$ (e.g., the wave number) represented through $h_{\rm X}$
is factored from the temperature and time dependence described by
$\sqrt{|\sigma|} g(t/t_{\sigma})$.

The decay of $\phi_{\rm X}(t)$ down from
the plateau $f_{\rm X}^{\rm c}$
is called the $\alpha$ process.
For this process, MCT predicts for $\sigma \to -0$
\begin{equation}
\phi_{\rm X}(t) = \tilde{\phi}_{\rm X}(t/t_{\sigma}^{\prime}),
\label{eq:superposition}
\end{equation}
which is also referred to as the superposition principle.
The temperature independent shape function
$\tilde{\phi}_{\rm X}(\tilde{t})$ -- referred to as
the $\alpha$-master function -- is to be evaluated
from the MCT equations at $T = T_{\rm c}$,
and the temperature dependence is given via the
time scale $t_{\sigma}^{\prime}$ introduced
in Eq.~(\ref{eq:t-sigma-prime}).
The initial part of the $\alpha$ process
is given by the von-Schweidler law,
$\phi_{\rm X}(t) =
f_{\rm X}^{\rm c} - h_{\rm X}
(t/t_{\sigma}^{\prime})^{b}$~\cite{Goetze91b}.
The superposition principle implies that the $\alpha$-relaxation
time $\tau_{\rm X}$ of any dynamical variable,
defined e.g. via the convention
$\phi_{\rm X}(\tau_{\rm X}) = 0.1$ which will be employed in this paper,
is proportional to $t_{\sigma}^{\prime}$, i.e.,
\begin{equation}
\tau_{\rm X} = C_{\rm X} \, t_{\sigma}^{\prime}.
\label{eq:tau-alpha-t-sigma-prime}
\end{equation}
This implies that the unspecified  microscopic time scale $t_{0}$,
which is common to all dynamical quantities,
can be eliminated by choosing a particular variable ${\rm X}$,
e.g., the collective density correlator $\phi(q,t)$ at the structure
factor peak position $q^{*}$,
and then plotting other quantities as a function of
$t / \tau_{q^{*}}$
where $\tau_{q^{*}}$ denotes the $\alpha$-relaxation time
of $\phi(q^{*},t)$.

For our model, the mentioned MCT universal predictions have been
successfully applied to analyze simulation data,
from which various characteristic quantities have been
extracted~\cite{BennemannBaschnagelPaul1999_incoherent,Aichele01-all,Baschnagel05}.
For example, the critical temperature $T_{\rm c}^{\rm MD} \approx 0.45$
was obtained from the consistent analysis of both
$\beta$ and $\alpha$ relaxations.
Critical nonergodicity parameters have been determined for
several correlators
by applying Eq.~(\ref{eq:factorization-theorem}) including
its leading correction~\cite{Franosch97-Fuchs98}
in the $\beta$ regime.
These simulation results can directly be compared with
our new first-principles theoretical calculations.

In addition, and more importantly,
our microscopic theory developed here
can make predictions concerning polymer specific features
which are outside the scope of the universal MCT predictions.
Through the comparative study of such first-principles theoretical
predictions and simulation results, we will propose
an atomistic interpretation of the slow structural and conformational
dynamics of supercooled polymer melts.

\section{Results and Discussions}
\label{sec:results}

\subsection{Collective structural dynamics}
\label{sec:results-2}

\begin{figure}[htb]
\includegraphics[width=0.5\linewidth]{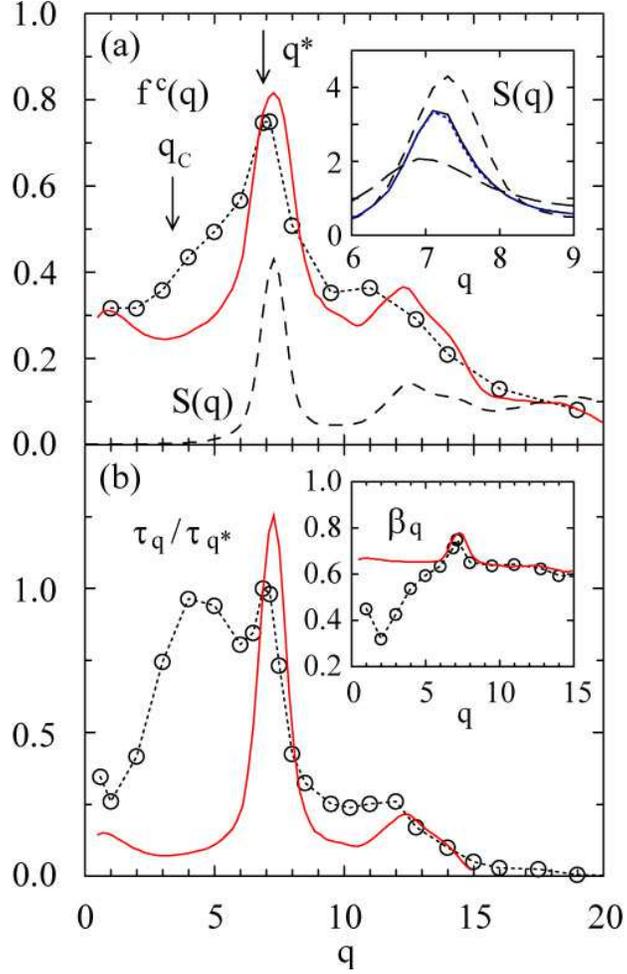}
\caption{
(Color online)
(a) Glass-form factors $f^{\rm c}(q)$ of the coherent density
correlators $\phi(q,t)$ versus $q$.
The circles represent the result from the simulation at
$T = 0.47$, and the solid line that from MCT.
The dashed line denotes the extrapolated $S(q)$ (multiplied by 0.1) at
$T_{\mathrm c}^{\rm MCT} \approx 0.277$.
The arrows indicate the peak positions $q^{*}$ and $q_{\rm C}$
of $S(q)$ and $S_{\rm C}(q)$
({\em cf.} Fig.~\ref{fig:Sq}).
The inset depicts the extrapolated $S(q)$ at $T_{\rm c}^{\rm MCT}$
(dashed line), and the simulated $S(q)$ at $T = 0.47$ (solid line),
0.48 (dotted line), and 1 (long-dashed line) around the peak $q^{*}$.
(b) Rescaled $\alpha$-relaxation times $\tau_{q}/\tau_{q^{*}}$
(main panel) and the stretching exponent $\beta_{q}$ (inset)
of $\phi(q,t)$ versus $q$.
The circles represent the result from the simulation at
$T = 0.47$, and the solid line that from MCT.}
\label{fig:fqc}
\end{figure}

We start by comparing theoretical and simulation results
for the critical glass-form factors $f^{\rm c}(q)$ of the coherent
density correlators $\phi(q,t)$.
The theoretical result for $f^{\rm c}(q)$ can be obtained
by solving Eq.~(\ref{eq:NEP}) provided static inputs at
$T_{\rm c}^{\rm MCT}$ are known.
($T_{\rm c}$ determined from the theory based on the analysis
of Eq.~(\ref{eq:NEP}) will be denoted as $T_{\rm c}^{\rm MCT}$
to discriminate it from $T_{\rm c}^{\rm MD}$ obtained from
previous analyses~\cite{Aichele01-all,Baschnagel05}
of the simulation data.)
If $T_{\mathrm c}^\mathrm{MCT}$ lies in the
range of temperatures for which simulation results are available,
the static inputs at $T_\mathrm{c}^\mathrm{MCT}$ can be determined
accurately.
However, a problem arises if $T_\mathrm{c}^\mathrm{MCT}$ is
found to be below the lowest
simulated $T$. This was the case in our study.
(See Sec.~\ref{sec:summary} for a possible explanation.)
So, we had to estimate the
required static input by a linear extrapolation
based on the simulated $S(q)$ at
$T = 0.47$ and 0.48.
(No extrapolation was necessary
for static intrachain correlation functions since they are
nearly independent of $T$ as shown in Ref.~\cite{Aichele04}.)
The dashed line in Fig~\ref{fig:fqc}(a) shows the extrapolated
$S(q)$ at $T_{\mathrm c}^\mathrm{MCT} \approx 0.277$
($< T_{\mathrm c}^\mathrm{MD} \approx 0.45$)
and the inset compares its behavior close to
the first-peak position $q^*$
with the simulated $S(q)$ at $T = 0.47$, 0.48
and 1. For the simulated $S(q)$ the peak height
increases and the peak position shifts to larger $q$ upon lowering
$T$ as discussed in Sec.~\ref{sec:static-1}.
The extrapolated $S(q)$ inherits this trend. This suggests that the
physics should not be significantly altered due to possible errors in our
extrapolation.

Figure~\ref{fig:fqc}(a) compares $f^{\mathrm c}(q)$ from MCT
with the simulation result determined in Ref.~\cite{Aichele01-all}.
For $q \gtrsim q^{*} = 6.9$,
i.e., for distances comparable to the average monomer separation,
we find a high degree of accord between theory and
simulation.
In particular, the agreement is quite good at $q^{*}$,
which is gratifying since the coherent dynamics for wave vectors
close to $q^{*}$ drives the glassy
slowing-down [{\em cf.} the second comment below Eq.~(\ref{eq:vertex})].
On the other hand, the theory fails to reproduce the shoulder
present in the simulation results at intermediate $q$ near the peak
$q_{\mathrm C} = 3.4$ of $S_{\rm C}(q)$.

A similar conclusion can be drawn for
the $\alpha$-relaxation time $\tau_{q}$,
defined via the convention $\phi(q,\tau_{q}) = 0.1$,
which is shown in Fig.~\ref{fig:fqc}(b).
$\tau_q$ depends nonmonotonically on $q$
and varies over an order of magnitude from $\tau_{q^*}\approx 1500$ to
$\tau_{q=16} \approx 40$ \cite{Aichele01-all}.
The theory semiquantitatively
captures these trends for $q \gtrsim q^*$, but
misses the peak around $q_\mathrm{C}$.
We also find corresponding deviations at $q \approx q_{\rm C}$
in the relaxation
stretching, as demonstrated in the inset of Fig.~\ref{fig:fqc}(b),
quantified in terms of the
stretching exponent $\beta_{q}$ which is obtained
via a Kohlrausch-law fit of the $\alpha$-decay part of the
correlator, $\phi(q,t) \propto
\exp[ - (t / \tau_{q}^{\prime})^{\beta_{q}}]$.

\begin{figure}[htb]
\includegraphics*[width=0.55\linewidth]{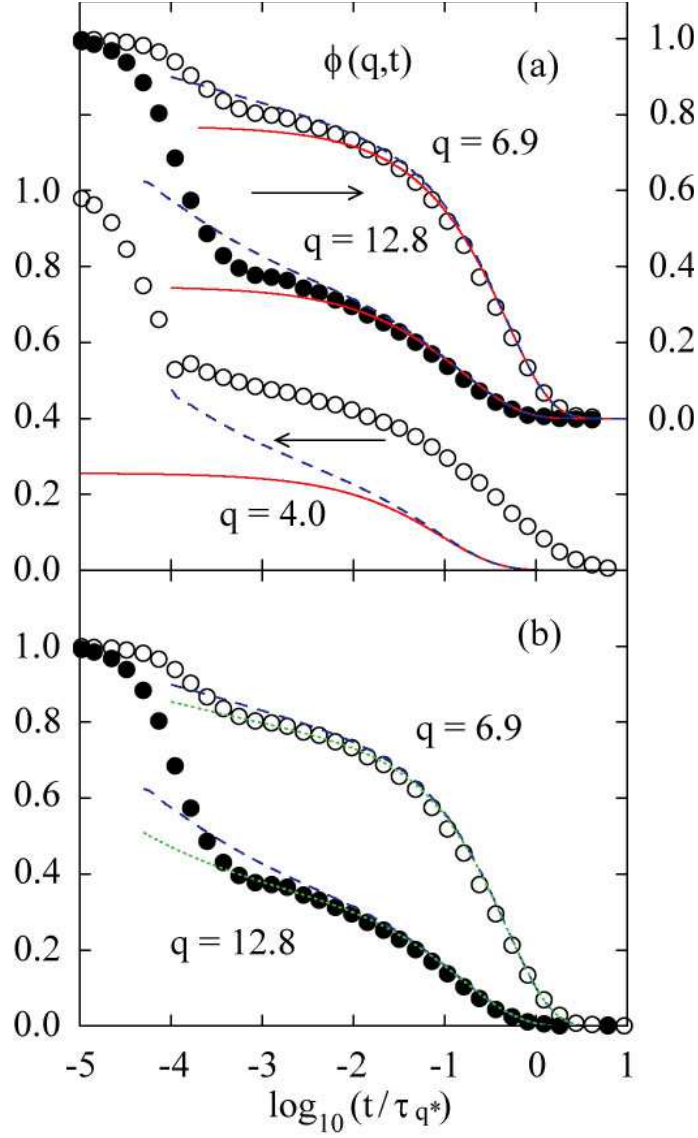}
\caption{
(Color online)
(a) $\phi(q,t)$ as a function of $t / \tau_{q^{*}}$ for $q = 4.0$
(left scale), 6.9 (right scale), and
12.8 (right scale).  $\tau_{q^*}$ is the
$\alpha$-relaxation time at $q^*$.
The circles refer to the simulation results
at $T = 0.47$, the solid lines to the MCT
$\alpha$-master curves, and the
dashed lines to the MCT curves at the distance
parameter $\epsilon^\mathrm{MCT} = -0.046$.
(b) $\phi(q,t)$ as a function of $t / \tau_{q^{*}}$ for $q = 6.9$
and 12.8.
The circles and the dashed lines are the same as in (a),
but here the dotted lines denoting the MCT curves
at the distance parameter $\epsilon^\mathrm{MCT} = -0.022$
are included as well.}
\label{fig:c-correlators}
\end{figure}

Circles in Fig.~\ref{fig:c-correlators}(a) shows $\phi(q,t)$ for $q
= 4.0$, 6.9, and 12.8, obtained from simulations at $T = 0.47$ which
is close to $T_{\rm c}^{\rm MD} \approx 0.45$. Clear evidence for
the presence of a two-step relaxation exists. As described in
Sec.~\ref{sec:theory-4}, MCT provides, up to a time scale $t_{0}$
common to all dynamical quantities, quantitative predictions for the
$\alpha$ regime in terms of the ($T$ independent) $\alpha$-master
curves, which are drawn as solid lines in
Fig.~\ref{fig:c-correlators}(a). The dynamics including the early
$\beta$ regime can be described by solving the MCT equations for a
temperature $T$ above $T_{\rm c}^{\rm MCT}$, which will be
characterized by the distance parameter $\epsilon^{\rm MCT} =
(T_{\rm c}^{\rm MCT} - T) / T_{\rm c}^{\rm MCT}$. The dashed curves
in Fig.~\ref{fig:c-correlators}(a) present such theoretical results
for $\epsilon^{\rm MCT} = -0.046$ which corresponds to the distance
between $T = 0.47$ and $T_{\mathrm c}^{\rm MD} \approx 0.45$ of the
simulation result. Since the separation parameter
controls the ratio of the time scales characterizing the $\beta$-
and $\alpha$-relaxation regimes [{\em cf.} Eqs.~(\ref{eq:t-sigma})
and (\ref{eq:t-sigma-prime})], we found that a better agreement in
the early $\beta$ regime can be achieved by treating $\epsilon^{\rm
MCT}$ as a fit parameter. The theoretical results for $\epsilon^{\rm
MCT} = -0.022$, which has been chosen so as to reproduce better the
time-scale ratio found in the simulation data, are shown as dotted
lines in Fig.~\ref{fig:c-correlators}(b). All the theoretical and
simulation curves shown in Figs.~\ref{fig:c-correlators}(a) and
\ref{fig:c-correlators}(b) are plotted versus $t / \tau_{q^{*}}$
with $\tau_{q^{*}}$ being the $\alpha$-relaxation time at $q^*$. In
this way, the unspecified time scale $t_{0}$ can be eliminated [{\em
cf.} discussion below Eq.~(\ref{eq:tau-alpha-t-sigma-prime})], and
the theoretical prediction on the $q$ dependence of the
$\alpha$-relaxation times can be examined.

Figure~\ref{fig:c-correlators}(a) and \ref{fig:c-correlators}(b)
demonstrate that, for $q = 6.9$ and 12.8, the MCT curves
quantitatively describe the simulation results in both the $\beta$ and
$\alpha$ regimes
including the relative magnitude of $\tau_{q}$
and the stretching of the relaxation.
(The disagreement for short times
arises mainly because,
as mentioned at the end of Sec.~\protect\ref{sec:theory-1},
the regular part of the memory kernel is not
included in our theory.)
On the other hand, the agreement is not satisfactory at
$q \approx q_{\mathrm C}$.
We thus conclude from Figs.~\ref{fig:fqc} and
\ref{fig:c-correlators} that,
except for a process that occurs at $q \approx q_{\mathrm C}$,
our theory describes the coherent structural dynamics of
polymer melts at a semiquantitative level.
Since, as will be discussed at the end of
Sec.~\ref{sec:results-3-3}, the dynamics at $q \approx q_{\rm C}$
does not appear to be directly related to the glass transition,
this verifies one of the principal predictions of our theory:
the emergence of the glassy slow dynamics is
connected to the increase of the first peak of $S(q)$,
i.e., to the local cage effect.
We will further comment on the dynamics at $q \approx q_{\rm C}$
below.

\subsection{Single-chain density correlators}
\label{sec:results-3-1}

We next consider the single-chain density correlators. This will be
done in terms of the averaged single
segment correlators $\phi^{\mathrm s}(q,t)$ and 
the (normalized) collective single chain correlators
$\phi^{\mathrm p}(q,t)$ defined by
\begin{eqnarray}
\phi^{\mathrm s}(q,t) &=&
(1/N) \sum_{a=1}^{N} F_{aa}^{\mathrm s}(q,t),
\\
\phi^{\mathrm p}(q,t) &=&
(1/N) \sum_{a,b=1}^{N} F_{ab}^{\mathrm s}(q,t) \, / \, w(q).
\label{eq:Fp-def}
\end{eqnarray}
$\phi^{\mathrm s}(q,t)$ is
defined solely in terms of the diagonal ($a=b$) elements of
$F_{ab}^{\mathrm s}(q,t)$,
i.e., it probes only the self-motion of monomers,
whereas
$\phi^{\mathrm p}(q,t)$ reflects also interference effects
from other monomers belonging to the same chain through the
off-diagonal ($a \ne b$) elements.
The simulation results for these correlators for our model
have been analyzed in Ref.~\cite{Aichele01-all}.

\begin{figure}[htb]
\includegraphics[width=0.5\linewidth]{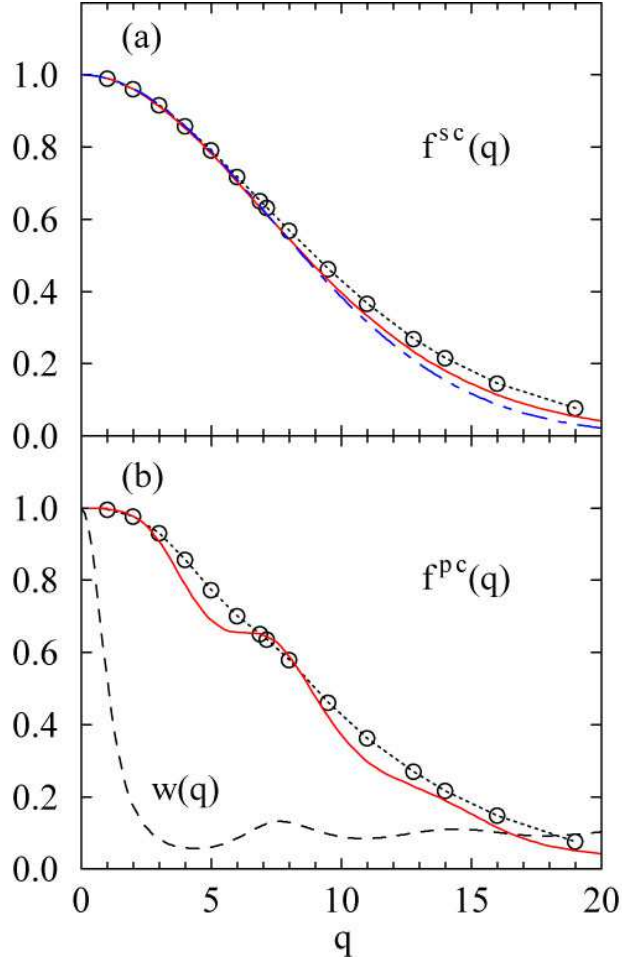}
\caption{
(Color online)
Glass-form factors
$f^{{\mathrm s} \, {\mathrm c}}(q)$ of the correlators
$\phi^{\mathrm s}(q,t)$ (a) and
$f^{{\mathrm p} \, {\mathrm c}}(q)$ of the correlators
$\phi^{\mathrm p}(q,t)$ (b) as a function of the wave number $q$.
The circles represent the result from the simulation at $T = 0.47$,
and the solid line that from MCT.
The dash-dotted line in (a) denotes
$f_{\rm G}^{{\mathrm s} \, {\mathrm c}}(q)$
based on the Gaussian approximation
(\ref{eq:fqs-Gaussian})
with the value $r_{\rm M}^{\rm c} = 0.098$
taken from the theoretical calculation.
The dashed line in (b) shows the simulated
$w(q)$ (multiplied by 0.1) at $T = 0.47$.}
\label{fig:Fqsc}
\end{figure}

\begin{figure}[htb]
\includegraphics*[width=0.5\linewidth]{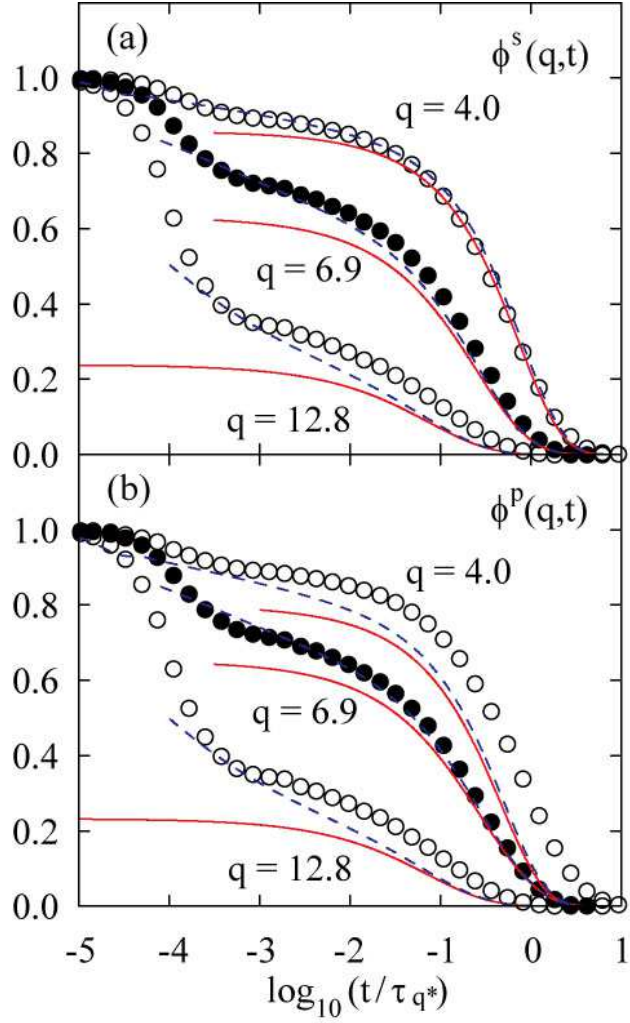}
\caption{
(Color online)
Single-chain density correlators
$\phi^{\mathrm s}(q,t)$ (a) and
$\phi^{\mathrm p}(q,t)$ (b)
as a function of $t / \tau_{q^{*}}$ for $q = 4.0$, 6.9, and 12.8.
$\tau_{q^*}$ is the $\alpha$-relaxation time of the coherent
density correlator $\phi(q,t)$ at $q = q^{*}$.
The circles refer to the simulation results at $T = 0.47$,
the solid lines to the MCT
$\alpha$-master curves, and the dashed lines to
the MCT curves at the distance
parameter $\epsilon^\mathrm{MCT} = -0.046$.}
\label{fig:incoherent}
\end{figure}

Theoretical and simulation results for the
critical nonergodicity parameters of
$\phi^{\mathrm s}(q,t)$ and $\phi^{\mathrm p}(q,t)$,
to be denoted as $f^{{\mathrm s} \, {\rm c}}(q)$ and
$f^{{\mathrm p} \, {\rm c}}(q)$,
are compared in Fig.~\ref{fig:Fqsc}(a) and \ref{fig:Fqsc}(b).
The theoretical results are determined from the equation
analogous to Eq.~(\ref{eq:NEP}), which can be obtained
by taking the $t \to \infty$ limit of
Eqs.~(\ref{eq:GLE-s}) and (\ref{eq:MCT-s}).
The full time-dependence of the correlators
$\phi^{\mathrm s}(q,t)$ and $\phi^{\mathrm p}(q,t)$
for representative wave numbers are shown in
Figs.~\ref{fig:incoherent}(a) and \ref{fig:incoherent}(b),
where the MCT $\alpha$-master curves and the
MCT curves at the distance parameter
$\epsilon^{\rm MCT} = -0.046$
are compared with the simulated correlators at $T = 0.47$.
All the curves in Figs.~\ref{fig:incoherent}(a) and
\ref{fig:incoherent}(b) are plotted versus $t / \tau_{q^{*}}$
with $\tau_{q^{*}}$ being the $\alpha$-relaxation time of the
coherent density correlator $\phi(q,t)$ at $q = q^{*}$.
In this way, the theoretical predictions
not only on the $q$ dependence of the $\alpha$-relaxation
times of $\phi^{\mathrm s}(q,t)$ and $\phi^{\mathrm p}(q,t)$, but also
on the relative time scale of the single-chain
and coherent dynamics can be tested.

We first notice from Fig.~\ref{fig:Fqsc}(a) that the theory (solid line)
describes well the simulated $f^{{\mathrm s} \, {\rm c}}(q)$ (circles)
including the wave numbers $q \lesssim q^{*}$.
Apparently, this seems inconsistent with the result for the
coherent dynamics where we found disagreement
at $q \approx q_{\rm C}$, but can be understood in the following way.
In the small-$q$ limit, there holds
$\phi^{\mathrm s}(q \to 0,t) = 1 - (q^{2}/6) g_{\rm M}(t) + O(q^{4})$
in terms of the monomer averaged MSD,
which implies the Gaussian approximation
$\phi_{\rm G}^{\mathrm s}(q,t) \approx \exp[- (q^{2}/6) g_{\rm M}(t)]$.
The discussion of $g_{\rm M}(t)$ will be presented in
Sec.~\ref{sec:results-3-3},
but let us mention here that the critical nonergodicity
parameter $6 (r_{\rm M}^{\rm c})^{2}$ of $g_{\rm M}(t)$ --
the plateau height of $g_{\rm M}(t)$ in the $\beta$ regime --
quantifies the size of the cage of a monomer
formed by its surroundings.
Thus, one expects the approximation
\begin{equation}
f_{\rm G}^{{\mathrm s} \, {\rm c}}(q) \approx \exp[ - q^{2} (r_{\rm M}^{\rm c})^{2} ],
\label{eq:fqs-Gaussian}
\end{equation}
to be valid at least for small $q$.
The dash-dotted line in Fig.~\ref{fig:Fqsc}(a) shows
$f_{\rm G}^{{\mathrm s} \, {\rm c}}(q)$
with the value $r_{\rm M}^{\rm c} = 0.098$
taken from the theoretical calculation.
It is seen that $f_{\rm G}^{{\mathrm s} \, {\rm c}}(q)$
describes well both the theoretical and simulated
$f^{{\mathrm s} \, {\rm c}}(q)$.
This reveals that the $q$-dependence of $f^{{\mathrm s} \, {\rm c}}(q)$
mostly reflects the spatial extent of the localized monomer motion
in the $\beta$-relaxation regime, and that
our theory quantitatively
describes the simulated $f^{{\mathrm s} \, {\rm c}}(q)$,
since it correctly predicts the size of the cage.
Figure~\ref{fig:incoherent}(a) furthermore indicates that,
not only the dynamics in the $\beta$-relaxation regime
where the correlators $\phi^{\mathrm s}(q,t)$ are close to the
plateaus $f^{{\mathrm s} \, {\rm c}}(q)$,
but also their $\alpha$ decay is well described by the
theory, including the relative time scale of the
single-chain and coherent dynamics.

In contrast to $f^{{\mathrm s} \, {\rm c}}(q)$,
we expect some interference effects due to chain connectivity
to be visible in
$f^{{\mathrm p} \, {\rm c}}(q)$
on top of the nearly Gaussian background just mentioned.
The theoretical prediction for $f^{{\mathrm p} \, {\rm c}}(q)$,
shown as the solid line in Fig.~\ref{fig:Fqsc}(b),
indeed exhibits an oscillatory feature which is in
phase with $w(q)$.
The oscillatory $q$ variation of $w(q)$ in turn reflects
the bonding effect as discussed in Ref.~\cite{Aichele04}.
In the simulated $f^{{\mathrm p} \, {\rm c}}(q)$,
shown as circles in Fig.~\ref{fig:Fqsc}(b),
the presence of such oscillations as predicted by the theory
is discernible,
though its amplitude is much smaller.
(The oscillation in the simulated $f^{{\mathrm p} \, {\rm c}}(q)$,
though tiny, can more easily be grasped in Fig.~8 of
the first article in Ref.~\cite{Aichele01-all} where
both of the simulated $f^{{\mathrm s} \, {\rm c}}(q)$ and
$f^{{\mathrm p} \, {\rm c}}(q)$ are plotted in one panel.)
This explains, e.g., why the theory does not describe so well
the simulated $\phi^{\mathrm p}(q,t)$ at $q=4.0$ [Fig.~\ref{fig:incoherent}(b)],
in spite of the fact that
it well describes the simulated $\phi^{\mathrm s}(q,t)$ at the same wave number
[Fig.~\ref{fig:incoherent}(a)]:
the oscillatory $q$ dependence of the plateau height and
the $\alpha$-relaxation time is more pronounced
in the theoretical prediction.

\subsection{Rouse-mode dynamics}
\label{sec:results-3-2}

\begin{figure}[htb]
\includegraphics[width=0.55\linewidth]{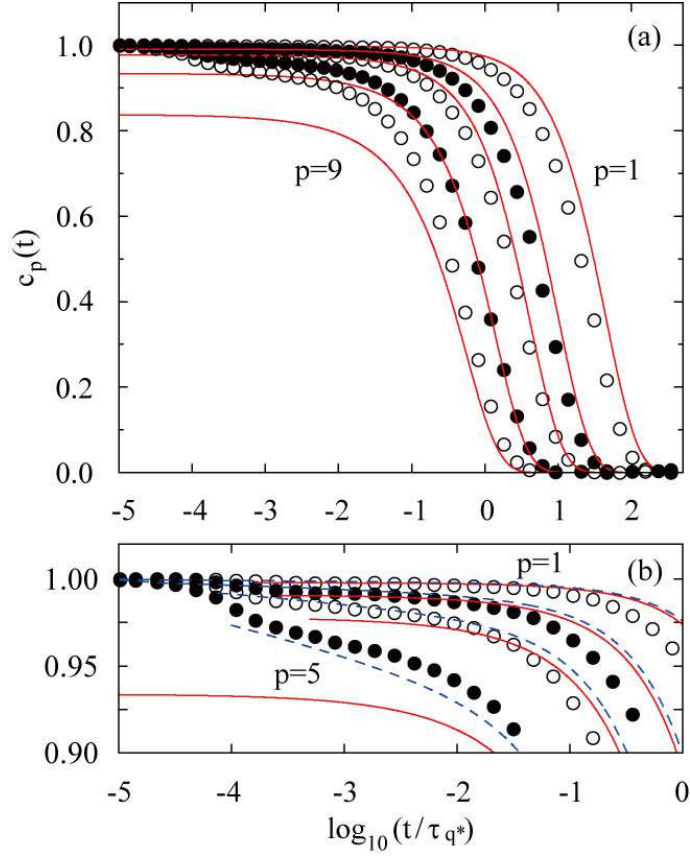}
\caption{
(Color online)
(a) Normalized Rouse-mode correlators
$c_{p}(t) = C_{pp}(t) / C_{pp}(0)$ as a
function of $t / \tau_{q^{*}}$
for $p=1$, 2, 3, 5, and 9 (from right to left).
$\tau_{q^*}$ is the $\alpha$-relaxation time of the coherent
density correlator $\phi(q,t)$ at $q = q^{*}$.
The circles refer to the simulation results at $T = 0.47$, and
the solid lines to the MCT $\alpha$-master curves.
(b) Enlargement of the $\beta$ region in (a);
the results for $p=9$ are omitted.
Here, dashed lines represent
the MCT curves at the distance
parameter $\epsilon^{\rm MCT} = -0.046$.}
\label{fig:Rouse-1}
\end{figure}

\begin{figure}[htb]
\includegraphics[width=0.5\linewidth]{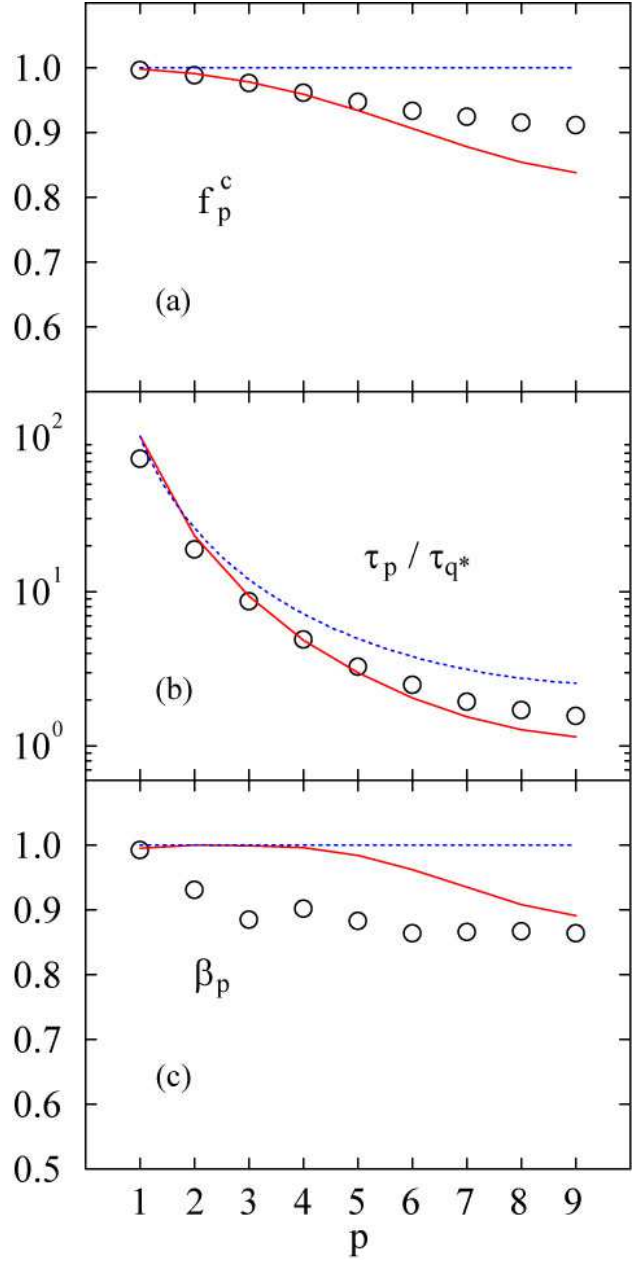}
\caption{
(Color online)
The plateau heights $f_{p}^{\mathrm c}$ (a),
the ratio $\tau_{p} / \tau_{q^{*}}$
of the $\alpha$-relaxation times (b),
and the stretching exponent $\beta_{p}$ (c)
of the Rouse-mode correlators $c_{p}(t)$
as a function of the mode index $p$.
The circles represent the result from the simulation at
$T = 0.47$, and the solid line that from MCT.
The dotted line in each panel
refers to pure Rouse behavior predicted by
our theory in the asymptotic limit of large $N$
({\em cf.} Appendix~\ref{appendix:Rouse-Fuchs}):
$f_{p}^{\rm c} = 1$,
$\tau_{p} \propto [\sin(p \pi/2N)]^{-2}$,
and $\beta_{p} = 1$.}
\label{fig:Rouse-2}
\end{figure}

We next turn our attention to the Rouse-mode correlators
$C_{pp'}(t)$ describing the chain conformational dynamics. 
Circles in Fig.~\ref{fig:Rouse-1}(a) show the
simulation results for the normalized Rouse-mode correlators
$c_{p}(t) = C_{pp}(t) / C_{pp}(0)$ at $T = 0.47$ for representative
mode indices $p$.
It is seen that $c_{p}(t)$ do not
clearly exhibit the two-step relaxation. This is because the
plateaus $f_{p}^{\rm c}$ of $c_{p}(t)$ are so large 
[$f_{p}^{\rm c} \gtrsim 0.9$ for simulation results as shown in
Fig.~\ref{fig:Rouse-2}(a)] that only about or less than 
10\% of the decay is left
for the relaxation towards the plateau. Thus, most of the relaxation
of $c_{p}(t)$ occurs in the $\alpha$ regime. We therefore included
for comparison only the MCT $\alpha$-master curves as solid lines in
Fig.~\ref{fig:Rouse-1}(a), and the comparison including the early
$\beta$ regime is done separately in Fig.~\ref{fig:Rouse-1}(b) which
highlights the plateau regime. For a more quantitative comparison of
the features in the $\alpha$ relaxation, theoretical and simulation
results for the plateau heights $f_{p}^{\rm c}$, the
$\alpha$-relaxation time $\tau_{p}$ defined via the convention
$c_{p}(\tau_{p}) = 0.1$, and the stretching exponent $\beta_{p}$
based on the Kohlrausch-law fit of $c_{p}(t)$ are compared in
Fig.~\ref{fig:Rouse-2} for all the Rouse modes $p$.

It is seen from Figs.~\ref{fig:Rouse-1} and \ref{fig:Rouse-2}
that our first-principles theory describes at a semiquantitative level
main features of the simulation results for $c_{p}(t)$,
such as the high plateau values $f_{p}^{\rm c}$
and the nearly exponential relaxation ($\beta_{p} \gtrsim 0.9$).
In particular, from the comparison of the ratio
$\tau_{p}/\tau_{q^{*}}$ of the $\alpha$-relaxation
times shown in Fig.~\ref{fig:Rouse-2}(b),
we see that
the theory provides a good description of the
time-scale separation
of the single-chain conformational fluctuations
(characterized by $\tau_{p}$)
from the local dynamics of the surrounding medium
($\tau_{q^{*}}$),
which becomes more pronounced with decreasing $p$.

The Rouse theory assumes a chain to be in a Markovian heat
bath, i.e., that all dynamical correlations in the
surroundings are much faster than the single-chain
dynamics~\cite{Doi86}.
However, since a polymer is surrounded by identical polymers,
the assumption of the time-scale separation cannot be justified
{\em a priori}.
Our microscopic theory developed here
verifies this central assumption from first principles:
our theory predicts the time-scale separation for small $p/N$,
and hence, the Markovian approximation can be justified
for small Rouse-mode indices (see Appendix~\ref{appendix:Rouse-Fuchs}).
The simulated results for the Rouse-mode correlators
do not exhibit pure Rouse behavior due to finite-$N$ effects
({\em cf.} Appendix~\ref{appendix:Rouse-Fuchs-4}),
and even such deviations
can be semiquantitatively accounted for by our theory
as demonstrated in Fig.~\ref{fig:Rouse-2} where
the pure Rouse behavior is also included for comparison. 

On the other hand, we also observe features predicted by
the Rouse model,
both in the simulated and theoretical results, 
already for such short chains as $N = 10$.
Let us recall from
Ref.~\cite{Bennemann99} that the matrix of the equilibrium values
of the (unnormalized) Rouse-mode correlators
for our model is close to diagonal, 
$C_{pp'}(t=0)\sim \delta_{pp'}$, and that
the amplitude $C_{pp}(t=0)$ for the smallest $p$ is
approximately given by the Gaussian result.
It is an important numerical observation within our
approach that the Rouse-mode correlators $C_{pp'}(t)$ remain
nearly diagonal for all the times, 
in agreement with the simulation result, and that
the time scale separation holds rather well
between the memory functions in Eq.~(\ref{eq:MCT-Rouse})
and the Rouse-mode correlators for the smallest $p$.
This holds because collective density fluctuations
at microscopic wave lengths dominate the
memory functions.
This is also the reason why the theory yields 
asymptotically in the large-$N$ limit
the Rouse model spectrum and the characteristic
$\sqrt{t}$ anomaly in the average segmental MSD 
(see Appendix~\ref{appendix:Rouse-Fuchs}).

\subsection{Mean-square displacements}
\label{sec:results-3-3}

\begin{figure}[htb]
\includegraphics*[width=0.55\linewidth]{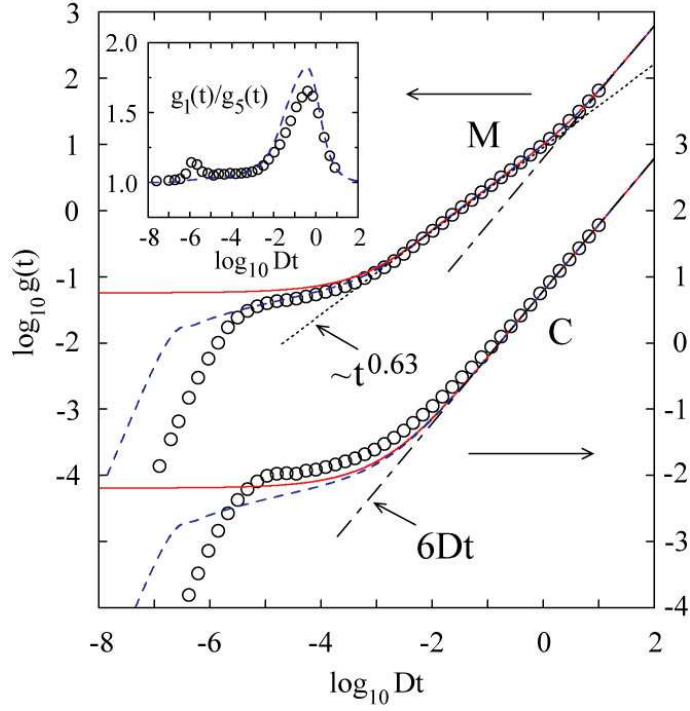}
\caption{
(Color online)
Double logarithmic presentation of the MSDs $g_{\mathrm M}(t)$
(labeled M, left scale)
and $g_{\mathrm C}(t)$ (labeled C, right scale) as a function of $Dt$.
The inset exhibits the ratio
$g_{1}(t) / g_{5}(t)$ (end- over middle-monomer MSD).
The circles refer to the simulation results at $T = 0.47$,
the solid lines to the MCT $\alpha$-master curves, and
the dashed lines to the MCT curves at the
distance parameter $\epsilon^\mathrm{MCT} = -0.046$.
The dash-dotted lines indicate the diffusion law, $6 Dt$, while
the dotted line shows the power law, $\sim t^{0.63}$.}
\label{fig:MSD-1}
\end{figure}

Now, let us see how the Rouse-mode dynamics affects
the single-chain diffusional processes.
Circles in Fig.~\ref{fig:MSD-1} shows the simulation
results for the monomer-averaged MSD,
$g_{\mathrm M}(t)$,
and the CM MSD, $g_{\mathrm C}(t)$.
The MSDs also exhibit a two-step relaxation:
after the short-time ballistic regime
$g_{\mathrm X}(t) \propto t^{2}$ (${\rm X} = {\rm M}$ or ${\rm C}$),
the increase of the MSD begins to be suppressed due to the
cage effect, and there appears the $\beta$ regime
where $g_{\rm X}(t)$ is close to a plateau
which will be denoted as $6 (r_{\rm X}^{\rm c})^{2}$.
The appearance of the plateau regime reflects the confined
dynamics of monomers inside the cage,
and the height of the plateau of the monomer MSD $g_{\rm M}(t)$
reflects the size of the cage.
We find both from theory and simulation
$r_{\rm M}^{c} \approx 0.1$ ({\em cf.} Fig.~\ref{fig:MSD-1}),
i.e., the amplitude of the confined dynamics inside the cage
is about 10\% of the monomer diameter.
The increase of $g_{\rm X}(t)$ above the plateau towards
the diffusion asymptote, $g_{\mathrm X}(t) = 6Dt$ with
$D$ denoting the diffusion constant,
is the $\alpha$ process of the MSD.
In contrast to $g_{\rm C}(t)$, $g_{\rm M}(t)$ in this regime
is significantly affected by chain connectivity
since the monomers participate in the conformational
motion and most of the conformational fluctuations
reflected by $c_{p}(t)$ occurs in the $\alpha$ regime
as mentioned above.
As a result, a polymer specific anomaly in the MSD
-- a subdiffusive ($\sim t^{x}$) regime -- emerges in
$g_{\mathrm M}(t)$ in the $\alpha$ regime.
Since MCT predicts the presence of the von-Schweidler-law process
as a universal feature ({\em cf.} Sec.~\ref{sec:theory-4}),
such a polymer-specific feature shows up
after the end of the von-Schweidler-law process
but before the onset of final diffusion~\cite{Chong02c}.

In Fig.~\ref{fig:MSD-1} the data are plotted versus $Dt$ so that the
simulated and theoretical curves coincide in the diffusive
late-$\alpha$ regime. This representation facilitates the comparison
in the $\beta$ and early-$\alpha$ regimes. (Plotting MSDs versus
$t/\tau_{q^{*}}$ as in Figs.~\protect\ref{fig:c-correlators} and
\protect\ref{fig:Rouse-1} leads to a horizontal shift of the
theoretical curves to the right by a factor of about 0.3 on the
\protect$\log_{\rm 10} t$ axis.) Figure~\ref{fig:MSD-1} demonstrates
that the theory in terms of the MCT $\alpha$-master curve (solid
line) describes the polymer-specific subdiffusive variation of the
$\alpha$-process, where $g_{\mathrm M}(t) \sim t^{x}$ with $x =
0.63$. Solving the MCT equations for $\epsilon^{\rm MCT} = -0.046$,
whose results are drawn as dashed lines in Fig.~\ref{fig:MSD-1}, the
description of the simulated $g_{\mathrm M}(t)$ can be extended to
about 7 decades in $t$. Our theory derives the Rouse result, $x =
1/2$, for $N \to \infty$ ({\em cf.}
Appendix~\ref{appendix:Rouse-Fuchs} and Ref.~\cite{Chong02c}), in
agreement with simulations of long chains in which topological
constraints are eliminated \cite{Shaffer94}. Thus, we suggest that
the somewhat larger $x = 0.63$ found in our $N=10$ model is a
deviation from the pure Rouse behavior due to finite-$N$ effects
({\em cf.} Appendix~\ref{appendix:Rouse-Fuchs-4}).

The inset of Fig.~\ref{fig:MSD-1} compares the ratio $g_{1}(t) /
g_{5}(t)$ of the MSDs for the end [$g_{1}(t)$] and central
[$g_{5}(t)$] monomers in a chain. This ratio is 1 in the ballistic
and diffusive regimes, and exhibits a maximum for times, where
conformational motion dominates the dynamics [i.e., for $t$ where
$g_{\mathrm M}(t) \sim t^{0.63}$]. The ratio is also close to 1 in
the $\beta$ regime, indicating that the cage effect slows the motion
of end and inner monomers in the same way. This is because the
dynamics in the $\beta$ regime is dominated by the confined dynamics
of the monomers inside the cage. The inset of Fig.~\ref{fig:MSD-1}
also reveals semiquantitative agreement between theory and
simulation for later times where the ratio exhibits a maximum. This
indicates that the motion of the end monomer and that of the central
monomer are well discriminated by the theory. Thus, chain-end
effects for the dynamics of a tagged chain are properly taken
into account by our theory, even though this effect was neglected in
the (static) direct correlation functions [{\em cf}.
Eq.~(\ref{eq:PRISM-c})]. This is because the matrix structure of
Eqs.~(\ref{eq:GLE-s}) and (\ref{eq:MCT-s}) is preserved for
describing the single-chain dynamics. The found agreement of the
ratio $g_{1}(t)/g_{5}(t)$ also shows that its maximum value, which
is somewhat smaller than 2 -- the result expected from the Rouse
theory -- reflects deviations due to finite-$N$ effects.

Concerning $g_{\mathrm C}(t)$, on the other hand, the theory is not so
satisfactory: besides the underestimated plateau height,
a careful examination of Fig.~\ref{fig:MSD-1}
indicates that the theoretical
$g_{\mathrm C}(t)$ enters the diffusion regime earlier
than the simulated one.
We will come back to this point in the following.
We only notice here that the disagreement in $g_{\rm C}(t)$
does not carry over to $g_{\mathrm M}(t)$,
because $g_{\mathrm M}(t) \gg g_{\mathrm C}(t)$
for times before the onset of the final diffusion regime.

\begin{figure}[htb]
\includegraphics*[width=0.55\linewidth]{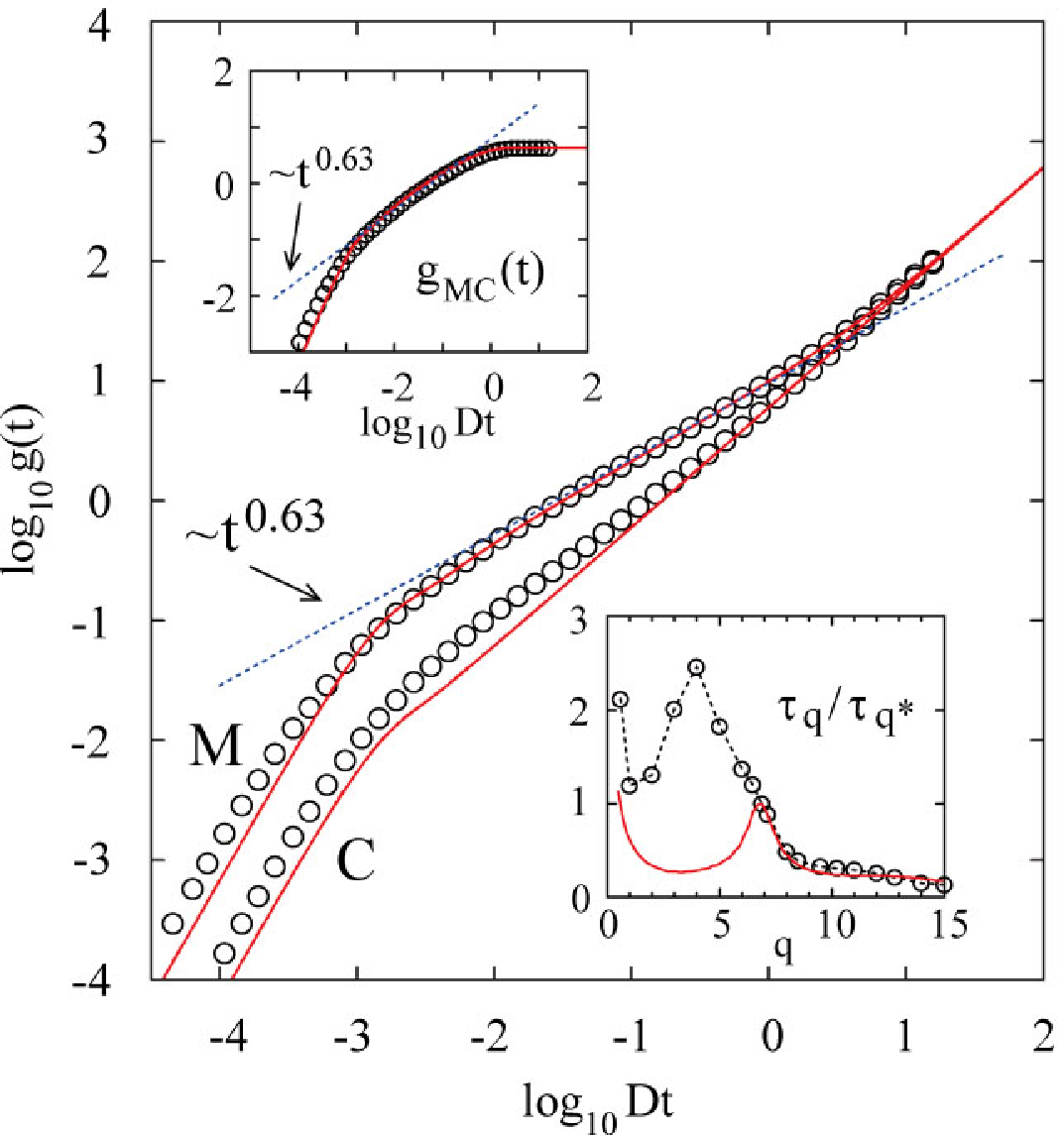}
\caption{
(Color online)
Double logarithmic presentation of $g_{\mathrm M}(t)$ (labeled M)
and $g_{\mathrm C}(t)$ (labeled C) versus $Dt$ at $T = 1$.
The upper inset exhibits $g_{\rm MC}(t) \equiv g_{\rm M}(t) - g_{\rm C}(t)$,
whereas the lower inset shows the $q$ dependence
the ratio $\tau_{q}/\tau_{q^{*}}$ of the $\alpha$-relaxation times
of the coherent density correlators $\phi(q,t)$ at $T = 1$.
The circles represent the result from the simulation,
and the solid line that from MCT.
The dotted line in the main panel and the upper inset
denotes the power law $\sim t^{0.63}$.}
\label{fig:MSD-2}
\end{figure}

As discussed above, our theory yields a subdiffusive, Rouse-like
behavior close to $T_\mathrm{c}$. Clearly, this polymer-specific
feature is also present in the simulation at high $T$. However, as
$T$ increases, the cage effect loses its importance, and it is thus
a priori not clear to what extent the theory can still be
applied. To examine this, we analyze in Fig.~\ref{fig:MSD-2} the
MSDs at $T = 1$, which is more than twice $T_{\mathrm
c}^\mathrm{MD}$. Here, the theory utilizes $S(q)$ taken directly
from the simulation at this temperature. Figure~\ref{fig:MSD-2}
indicates that, beyond the short-time regime, the agreement between
theory and simulation is very good for $g_{\mathrm M}(t)$. In
particular, we find $g_\mathrm{M}(t) \sim t^{0.63}$ with the same
exponent. Thus, though originally developed to describe glassy
dynamics, our theory can also properly deal with the conformational
dynamics in normal liquid states.

The upper inset of Fig.~\ref{fig:MSD-2} exhibits the function
$g_{\rm MC}(t) \equiv g_{\rm M}(t) - g_{\rm C}(t)$.
According to Eq.~(\ref{eq:MSD-monomer-averaged}),
this function highlights the contributions from the chain conformational
fluctuations to $g_{\rm M}(t)$, and depends only on the Rouse modes
of nonzero mode indices, $p>1$.
The inset clearly indicates that the subdiffusive behavior
$\sim t^{0.63}$ entirely comes from the Rouse modes,
and that the exponent 0.63 does not reflect a crossover effect from the
pure Rouse behavior to the final diffusion
(i.e., $A t^{0.5} + 6 Dt \nsim t^{0.63}$),
but is indeed due to finite-$N$ effects.

On the other hand, we find at $T = 1$ again the same disagreement for
$g_{\mathrm C}(t)$ and for the ratio
$\tau_q/\tau_{q^*}$:
(i) The theoretical
$g_{\mathrm C}(t)$ enters the diffusion regime earlier than the simulated
one which additionally exhibits a subdiffusive behavior, $g_{\mathrm C}(t)
\sim t^y$ with $y \approx 0.8$,
known as ``anomalous CM diffusion''~\cite{Paul04,Guenza02}.
(ii) The theoretical $\alpha$-relaxation times $\tau_q$ of
$\phi(q,t)$ agree
quantitatively with the simulation for
$q \gtrsim q^{*}$, but not for $q \approx q_{\mathrm C}$
({\em cf.} the lower inset of Fig.~\ref{fig:MSD-2}).
Thus, the disagreements observed at $T=0.47$
are already present at high $T$, suggesting that
they are not directly related to the glass transition.

\section{Summary and concluding remarks}
\label{sec:summary}

In this paper, we proposed a unified first-principles description of
the collective structural slowing-down and of the single-chain
conformational fluctuations in a melt of unentangled polymers.
The description requires static input which can be
taken directly from simulations, uses approximations like the
equivalent-site approximation that
can be tested explicitly~\cite{Aichele04}, and attains
semiquantitative agreement with simulation results
concerning collective as well as single-chain dynamics.
Our comparative study of theoretical predictions and
simulation data identifies local
structural correlations of monomers as the origin for the
onset of glassy slow dynamics.
It is also shown that the chain connectivity causes the
polymer-specific long-time anomalies of the $\alpha$ process,
which manifest themselves in the subdiffusive monomer
mean-square displacement.
Thus, the widely used picture of
polymer transport in unentangled melts,
the Rouse model -- including deviations
due to finite $N$ -- emerges from our first-principles
approach ({\em cf.} Appendix~\ref{appendix:Rouse-Fuchs}).

On the other hand, we also found deviations between theoretical
and simulation results.
Though probably not directly related to the glass transition
({\em cf.} the end of Sec.~\ref{sec:results-3-3}),
the most noticeable disagreement occurs in the
collective density fluctuations on the length scale of a chain,
$q \approx q_{\rm C}$ (Fig.~\ref{fig:fqc}).
This disagreement might explain why
our theory underestimates $T_{\rm c}$
($T_{\rm c}^{\rm MCT} \approx 0.277 <
T_{\rm c}^{\rm MD} \approx 0.45$).
There are discernible slow modes at $q \approx q_{\rm C}$,
not accounted for by our theory,
which appear to couple to the relaxation at other wave vectors.
The only way in which our theory can compensate this additional
coupling is by making the cage effect stronger,
i.e., by increasing the first peak of $S(q)$ via a decrease
of $T_{\rm c}^{\rm MCT}$.
Another possible source for the underestimation of
$T_{\mathrm c}^{\rm MCT}$ could result from
the neglect of the triple direct correlation
functions $c_{3}$ in our theoretical calculations.
It was found from an MCT analysis for a model of orthoterphenyl
that including $c_{3}$ considerably increases
$T_{\rm c}^{\rm MCT}$~\cite{Rinaldi01}.
For our polymer model, $c_{3}$
has been determined from simulations in Ref.~\cite{Aichele04},
and in principle
it would be rather straightforward to take them into account
in our theory. Unfortunately,
the statistical accuracy of the simulated $c_{3}$ was not sufficient
to allow for a meaningful test to investigate
to what extent the inclusion of $c_{3}$ affects
the value of $T_{\rm c}^{\rm MCT}$.

Inspection of Fig.~\ref{fig:MSD-2} implies that the
disagreement found in the theoretical predictions --
the one in the collective dynamics at $q \approx q_{\rm C}$
and the other in $g_{\rm C}(t)$ concerning the
anomalous CM dynamics --
might be somehow related,
since these are the features for which our theory
does not work well.
Superficially, this conjecture agrees with the physics
discussed in Ref.~\cite{Guenza02}.  There,
the anomalous CM MSD is connected to the polymer coils interacting as
spheres of radius of gyration $R_\mathrm{g}$, and the dynamics at
$q_{\mathrm C} \approx 2 \pi / R_\mathrm{g}$ reflects the polymer
packing.
This implies that taking into account the spatial correlation
of CMs through the CM structure factor $S_{\rm C}(q)$
might improve the theoretical results on
$\phi(q,t)$ at $q \approx q_{\rm C}$ and $g_{\rm C}(t)$.
Implementing this idea is rather
straightforward ({\em cf}. Ref.~\cite{Chong04} for a related problem).
However, no improvement was obtained in our case, certainly
because $S_{\rm C}(q)$ at $q_{\mathrm C}$ is already close to 1
({\em cf}. the inset of Fig.~\ref{fig:Sq}).
Thus, the static coupling between the CMs in our model is very weak.
Furthermore, also their dynamic coupling is found to be weak,
as evidenced by the close agreement of
their coherent and incoherent intermediate scattering functions at
$q_{\mathrm C}$~\cite{Aichele-PhD}.  At present, it is not clear how to
improve the theory to account for the deviations observed
in the collective dynamics at $q \approx q_{\mathrm C}$ and
in the CM MSD $g_{\rm C}(t)$.
It would be interesting to investigate
to what extent such features are universal or
model dependent. Only comparison with other models 
can elucidate this point.

There is another interesting related issue concerning the
collective dynamics at $q \approx q_{\rm C}$.
One observes from comparing Fig.~\ref{fig:fqc}(b)
with the lower inset of Fig.~\ref{fig:MSD-2} that 
the simulation result for the ratio
$\tau_{q_{\rm C}} / \tau_{q^{*}}$
of the $\alpha$-relaxation time at $q_{\rm C}$ to the
one at the structure factor peak position $q^{*}$
decreases with decreasing $T$ toward $T_{\rm c}^{\rm MD}$.
A similar feature was observed in a simulation result for
a model of orthoterphenyl which also exhibits some unusual
properties at intermediate wave numbers corresponding to
$q_{\rm C}$ of the present polymer model~\cite{Chong04}.
A similar $T$ dependence of the ratio $\tau_{q} / \tau_{q^{*}}$
at intermediate $q$ range ($\approx 0.4 q^{*}$) was also
found in the coherent neutron-scattering results
for a real polymer system~\cite{Farago02}.
As discussed in some detail in Ref.~\cite{Chong04},
such $T$ dependence of the ratio of the $\alpha$-relaxation times
is beyond the implication of MCT.
Thus, further investigations are necessary for a comprehensive
understanding of the yet theoretically unexplained dynamics
at intermediate wave numbers
which are observable in simulation and experimental
data for polymer systems.

\begin{acknowledgments}
Financial support by
the DFG and MENRT (IRTG ``Soft Matter''),
the IUF,
the DAAD (Grant No. D/00/07994),
the ESF SUPERNET Programme, and
Grant-in-Aids for scientific
research from the
Ministry of Education, Culture, Sports, Science and
Technology of Japan (Grant No. 17740282)
is gratefully acknowledged.
\end{acknowledgments}

\appendix

\section{Derivation of the MCT equations of motion}
\label{appendix:MCT}

This appendix is devoted to the derivation of the MCT
equations of motion for general flexible (in the sense
that constituent atoms are bonded by some non-rigid potential)
molecules.
Additional approximations,
introduced specifically for handling polymeric systems, are
discussed in the main text.

\subsection{Zwanzig-Mori equation of motion}
\label{appendix:MCT-1}

We start from the derivation of an exact equation of motion
for the site-site density correlators
$F_{ab}(q,t) =
\langle \rho_{a}({\bf q})^{*} e^{i {\cal L}t}
\rho_{b}({\bf q}) \rangle / n$
based on the Zwanzig-Mori projection-operator
formalism~\cite{Hansen86}.
Here, ${\cal L}$ denotes the Liouville operator
\begin{equation}
i{\cal L} =
\sum_{i=1}^{n} \sum_{a=1}^{N}
 {\bf v}_{i}^{a} \cdot \frac{\partial}{\partial {\bf r}_{i}^{a}} -
\frac{1}{m} \sum_{i,j=1}^{n} \sum_{a,b=1}^{N}
\frac{\partial U(| {\bf r}_{i}^{a} - {\bf r}_{j}^{b} |)}
{\partial {\bf r}_{i}^{a}}
\cdot \frac{\partial}{\partial {\bf v}_{i}^{a}},
\end{equation}
where ${\bf r}_{i}^{a}$ (${\bf v}_{i}^{a}$)
denotes the position (velocity) of the
site $a$ in the $i$th molecule.
The interaction potential $U(r)$ comprises both the
intra- and inter-molecular contributions
({\em cf.} Sec.~\ref{sec:model}).

To derive the exact equation for $F_{ab}(q,t)$,
let us also introduce the longitudinal current density
fluctuations
$j_{a}({\bf q}) = \sum_{i=1}^{n}
v_{i,z}^{a} \, e^{ i {\bf q} \cdot {\bf r}_{i}^{\, a} }$
where the wave vector
${\bf q}$ is chosen along the $z$ axis,
and $v_{i,z}^{a}(t)$ denotes the $z$ component of the
velocity.
$\rho_{a}({\bf q})$ and $j_{a}({\bf q})$
satisfy the continuity equation
\begin{equation}
\dot{\rho}_{a}({\bf q}) = i {\cal L} \rho_a({\bf q}) = i q \,
j_{a}({\bf q}), 
\label{eq:conti}
\end{equation}
in which the dot denotes the time derivative.
For flexible-molecule systems whose kinetic energy reads
$\sum_{i,a} m({\bf v}_{i}^{ \, a})^{2} / 2$,
the static longitudinal current
correlation function is given by
\begin{equation}
J_{ab}(q) = \frac{1}{n}
\langle
j_{a}({\bf q})^{*}
j_{b}({\bf q})
\rangle = \delta_{ab} \, v^{2},
\label{eq:J-def}
\end{equation}
due to the equipartition theorem.
Unlike for rigid molecules~\cite{Chong98},
there is no off-diagonal
element and no wave-number dependence in $J_{ab}(q)$.

Let us introduce two row vectors
$\mbox{\boldmath $\rho$}({\bf q})$ and ${\bf j}({\bf q})$,
whose components are $\rho_{a}({\bf q})$ and $j_{a}({\bf q})$,
respectively.
Combining $\mbox{\boldmath $\rho$}({\bf q})$ and ${\bf j}({\bf q})$
to form a new row vector
${\bf C}({\bf q}) \equiv
( \,\,
\mbox{\boldmath $\rho$} ({\bf q}) \,\,\,\,
{\bf j}({\bf q}) \,\, )$,
we introduce the following projection operator ${\cal P}$
which acts on some row vector ${\bf X}({\bf q})$:
\begin{equation}
{\cal P} \, {\bf X} \equiv
{\bf C} ({\bf C}, {\bf C})^{-1} ({\bf C}, {\bf X}).
\end{equation}
Here, the inner product of two row vectors,
${\bf A}_{1}$ and ${\bf A}_{2}$,
shall be defined as the
canonical ensemble average,
$({\bf A}_{1}, {\bf A}_{2}) \equiv
\langle {\bf A}_{1}^{\dagger} {\bf A}_{2}
\rangle / n$,
in which ${\bf A}_{1}^{\dagger}$ denotes a column vector
adjoint to ${\bf A}_{1}$, and
the factor $1/n$ is a matter of convention.
The matrix $({\bf C}, {\bf C})$ can thus be expressed in terms of
the site-site static correlation functions as
\begin{equation}
({\bf C}, {\bf C}) =
\left(
\begin{array}{cc}
{\bf S}(q) & {\bf 0}    \\
{\bf 0}    & {\bf J}(q)
\end{array}
\right) \, ,
\end{equation}
and its inverse is trivially given in terms of
${\bf S}^{-1}(q)$ and ${\bf J}^{-1}(q)$. 

With the projection operator ${\cal P}$ so defined,
the standard procedure of the Zwanzig-Mori formalism
leads to the following equation of motion for $F_{ab}(q,t)$:
\begin{equation}
\ddot{F}_{ab}(q,t) +
\sum_{x=1}^{N} \Omega_{ax}^{2}(q) \, F_{xb}(q,t) +
\sum_{x=1}^{N} \int_{0}^{t} dt' \, M_{ax}(q,t-t') \, \dot{F}_{xb}(q,t') = 0.
\label{eq:GLE-a}
\end{equation}
Here the characteristic frequency matrix reads
\begin{equation}
\Omega_{ab}^{2}(q) = q^{2} v^{2} S^{-1}_{ab}(q),
\label{eq:GLE-b}
\end{equation}
and the formal expression for the site-site memory kernel
$M_{ab}(q,t)$ is given by
\begin{equation}
M_{ab}(q,t) =
\frac{1}{nv^{2}}
\langle f_{a}({\bf q})^{*}
\exp (i {\cal QLQ} t)
f_{b}({\bf q}) \rangle,
\label{eq:memory3}
\end{equation}
in terms of the fluctuating random force
\begin{equation}
f_{a}({\bf q}) =
\dot{j}_{a}({\bf q})
- i q v^{2}
\sum_{\nu} \rho_{\nu}({\bf q}) S_{\nu a}^{-1}(q),
\end{equation}
which evolves with the generator ${\cal QLQ}$, where ${\cal Q}
\equiv 1 - {\cal P}$. So far, no approximation has been invoked, and
the above equation for $F_{ab}(q,t)$ is formally exact.

\subsection{Mode-coupling approximation}
\label{appendix:MCT-2}

The basic idea behind the mode-coupling theory is that the
fluctuation of a given dynamical variable decays,
at intermediate and long times, predominantly
into pairs of hydrodynamic modes associated with
quasi-conserved dynamical variables.
It is reasonable to expect that the decay of the memory
function at intermediate and long times is dominated
by those mode correlations which have the longest
relaxation times.
The sluggishness of the structural relaxation processes
in glass-forming systems suggests that the slow decay
of the memory function at long times is basically due to
couplings to wave-vector-dependent pair density modes of the form
$A_{\lambda \mu}({\bf k},{\bf p}) \equiv
\rho_{\lambda}({\bf k})
\rho_{\mu}({\bf p})$.
The simplest way to extract such slowly-decaying part
is to introduce another projection
operator ${\cal P}_{2}$ which projects any variable
onto the subspace spanned by $A_{\lambda \mu}({\bf k},{\bf p})$.
Translational invariance of the system
implies that the only $A_{\lambda \mu}({\bf k},{\bf p})$,
whose inner products with
a dynamical variable $X({\bf q})$ are non-zero,
are for the wave vectors ${\bf p}$
satisfying ${\bf p} = {\bf q} - {\bf k}$.
From here on, we denote by $A_{\lambda \mu}$ those
$A_{\lambda \mu}({\bf k}, {\bf p})$
in which ${\bf p} = {\bf q} - {\bf k}$, and we define
\begin{equation}
{\cal P}_{2} X \equiv
\frac{1}{2}
\sum_{\bf k}
\sum_{\lambda, \mu, \lambda', \mu'}
A_{\lambda \mu}
(A_{\lambda \mu}, A_{\lambda' \mu'})^{-1}
(A_{\lambda' \mu'}, X).
\label{eq:P2def}
\end{equation}
Here the factor $1/2$ is to avoid
the double counting in the summation over the wave vectors, and
the inverse is defined via
\begin{equation}
\sum_{\lambda', \mu'}
(A_{\lambda \mu}, A_{\lambda' \mu'})
(A_{\lambda' \mu'}, A_{\lambda'' \mu''})^{-1} =
\delta_{\lambda \lambda''} \delta_{\mu \mu''}.
\label{eq:inv2def}
\end{equation}
It is readily verified that ${\cal P}_{2}$
is idempotent and Hermitian.

The first approximation in the mode-coupling approach thus
corresponds to replacing the time-evolution operator
$\exp (i {\cal QLQ} t)$ by its projection on the subspace
spanned by $A_{\lambda \mu}$:
$\exp (i {\cal QLQ} t)
\approx
{\cal P}_{2} \exp (i {\cal QLQ} t) {\cal P}_{2}$.
Under this approximation, the memory function reads
\begin{equation}
M_{ab}(q,t) = \frac{1}{nv^{2}} \,
\langle \, {\cal P}_{2} f_{a}({\bf q})^{*} \,
\exp (i {\cal QLQ} t) \,
{\cal P}_{2} f_{b}({\bf q}) \, \rangle.
\label{eq:KLMCT}
\end{equation}
The second approximation is to factorize averages of products,
evolving in time with the generator ${\cal QLQ}$,
into products of averages formed with variables
evolving with ${\cal L}$ (factorization approximation):
\begin{equation}
\frac{1}{n^{2}}
\langle
\rho_{\lambda}({\bf k})^{*}
\rho_{\mu}({\bf p})^{*}
e^{i {\cal QLQ} t}
\rho_{\lambda'}({\bf k})
\rho_{\mu'}({\bf p}) \rangle
\approx
F_{\lambda \lambda'}(k,t) \, F_{\mu \mu'}(p,t).
\label{eq:factori-appendix}
\end{equation}
Specializing this approximation to $t=0$,
it follows from Eq.~(\ref{eq:inv2def}) that
the denominator in Eq.~(\ref{eq:P2def}) is given by
\begin{equation}
(A_{\lambda \mu}, A_{\lambda' \mu'})^{-1} =
\frac{1}{n} \,
S_{\lambda \lambda'}^{-1}(k) \,
S_{\mu \mu'}^{-1}(p).
\label{eq:inv2}
\end{equation}

Let us obtain the explicit expression for the
projected random force,
\begin{equation}
{\cal P}_{2} f_{a}({\bf q}) =
{\cal P}_{2} \dot{j}_{a}({\bf q})
- i q v^{2}
\sum_{\nu}
({\cal P}_{2} \rho_{\nu}({\bf q})) S_{\nu a}^{-1}(q).
\label{eq:P2rangen}
\end{equation}
To this end, we need to evaluate triple correlations
$(A_{\lambda \mu}, \dot{j}_{a}({\bf q}))$ and
$(A_{\lambda \mu}, \rho_{\nu}({\bf q}))$.
The former can be expressed as
\begin{equation}
(\rho_{\lambda}({\bf k})
\rho_{\mu}({\bf p}),
\dot{j}_{a}({\bf q})) =
i k_{z} \frac{1}{n}
\langle
j_{\lambda}^{*}({\bf k})
\rho_{\mu}^{*}({\bf p})
j_{a}({\bf q})
\rangle
+ i p_{z} \frac{1}{n}
\langle
\rho_{\lambda}^{*}({\bf k})
j_{\mu}^{*}({\bf p})
j_{a}({\bf q})
\rangle,
\end{equation}
where we have used the relation
$\langle A \dot{B} \rangle = - \langle \dot{A} B \rangle$
and the continuity equation~(\ref{eq:conti}).
$k_{z}$ ($p_{z}$) denotes the $z$ component of the
vector ${\bf k}$ (${\bf p}$).
Since
$\langle v_{i,z}^{a} v_{j,z}^{b} \rangle = \delta_{ij} \delta_{ab} v^{2}$,
there hold
$\langle
j_{\lambda}^{*}({\bf k})
\rho_{\mu}^{*}({\bf p})
j_{a}({\bf q})
\rangle /n =
\delta_{a \lambda} v^{2} S_{\lambda \mu}(p)$ and
$\langle
\rho_{\lambda}^{*}({\bf k})
j_{\mu}^{*}({\bf p})
j_{a}({\bf q})
\rangle / n =
\delta_{a \mu} v^{2} S_{\lambda \mu}(k)$, leading to
\begin{equation}
(\rho_{\lambda}({\bf k})
\rho_{\mu}({\bf p}),
\dot{j}_{a}({\bf q})) =
i k_{z}
\delta_{a \lambda} v^{2} \, S_{\lambda \mu}(p) +
i p_{z}
\delta_{a \mu} v^{2} \, S_{\lambda \mu}(k).
\end{equation}
The other triple correlation
can be expressed in terms of the three-site static structure factor:
\begin{equation}
(\rho_{\lambda}({\bf k}) \rho_{\mu}({\bf p}),
\rho_{\nu}({\bf q})) =
\frac{1}{n}
\langle
\rho_{\lambda}({\bf k})^{*} \rho_{\mu}({\bf p})^{*}
\rho_{\nu}({\bf q})
\rangle \equiv
S_{\lambda \mu \nu}({\bf k}, {\bf p}, {\bf q}).
\end{equation}
In the present study, the convolution approximation
developed in Ref.~\cite{Chong98} shall be employed:
\begin{equation}
S_{\lambda \mu \nu}({\bf k}, {\bf p}, {\bf q}) \approx
\sum_{\sigma} S_{\lambda \sigma}(k) \, S_{\mu \sigma}(p) \,
S_{\nu \sigma}(q).
\label{eq:convolution}
\end{equation}
Using these results, we finally obtain
\begin{eqnarray}
{\cal P}_{2} f_{a}({\bf q}) &=&
- \frac{i v^{2}}{n}
\sum_{\bf k}
\sum_{\lambda}
k_{z} \, [\, \delta_{\lambda a} - S_{\lambda a}^{-1}(k) \, ] \,
\rho_{\lambda}({\bf k})
\rho_{a}({\bf p})
\nonumber \\
&=&
- \frac{i \rho v^{2}}{n}
\sum_{\bf k}
\sum_{\lambda}
k_{z} [ \, c_{\lambda a}^{\rm intra}(k) + c_{\lambda a}(k) \, ] \,
\rho_{\lambda}({\bf k})
\rho_{a}({\bf p}).
\label{eq:P2ranall}
\end{eqnarray}
Here, $c_{ab}^{\rm intra}(q) = [\delta_{ab} - w_{ab}^{-1}(q)]/\rho$ denotes
the {\em intra-molecular} direct correlation
function~\cite{Chandler86},
whereas $c_{ab}(q)$ is the {\em inter-molecular} one
defined in Eq.~(\ref{eq:c-def}).
Thus, the projected random force naturally comprises the
intra-molecular as well as inter-molecular contributions.

The MCT expression for $M_{ab}(q,t)$ can be obtained by substituting
Eq.~(\ref{eq:P2ranall}) into Eq.~(\ref{eq:KLMCT}) and then invoking
the factorization approximation~(\ref{eq:factori-appendix}), but let
us make here a comment on the intra-molecular contribution. Within
the same approach outlined above, one can derive the MCT expression
for the memory kernel for the site-site {\em transversal} current
density correlator, whose $q \to 0$ limit is related to the
shear-stress autocorrelation function $G(t)$~\cite{Hansen86}. It is
well known for unentangled polymer chains that $G(t)$ exhibits a
power-law decay $\sim t^{-1/2}$ for long times. According to the
Rouse theory, this polymer-specific decay in $G(t)$ is accounted for
by the intrachain (or Rouse-mode) contributions~\cite{Doi86}.
However, we found that our intra-molecular contributions given in
terms of $c^{\rm intra}$ do not lead to this Rouse model result for
$G(t)$. This implies that a completely different kind of approach is
necessary for a proper treatment of the intra-molecular
contributions in the coherent moduli. Indeed, we found a
reasonable approach starting from a different projection operator
for the intra-molecular forces, which reproduces the Rouse model
result for $G(t)$. This issue, however, shall not be investigated
further here, and will be studied in a forthcoming article. Let us
only mention that (i) even with the inclusion of such
intra-molecular contributions to the random force or to the memory
kernel, we confirmed that all the theoretical results presented in
the main text are not much affected, and (ii) neglecting
intra-molecular contributions to the fluctuating force does not mean
that intra-molecular couplings are completely discarded in our
theory, since the intrachain static correlations are properly taken
into account via $w_{ab}(q)$.

The following expression for
${\cal P}_{2} f_{a}({\bf q})$ shall therefore be
employed in the present work:
\begin{equation}
{\cal P}_{2} f_{a}({\bf q}) =
- \frac{i \rho v^{2}}{n}
\sum_{\bf k}
\sum_{\lambda}
(\widehat{\bf q} \cdot {\bf k}) c_{\lambda a}(k) \,
\rho_{\lambda}({\bf k})
\rho_{a}({\bf p}),
\label{eq:MCT-random-force}
\end{equation}
in which we have expressed $k_{z}$ as
$\widehat{\bf q} \cdot {\bf k}$ with
$\widehat{\bf q} = {\bf q} / q$. 
With the use of the factorization
approximation~(\ref{eq:factori-appendix}),
this leads to
\begin{eqnarray}
M_{ab}(q,t) &=&
\frac{\rho v^{2}}{(2 \pi)^{3}}
\sum_{\lambda,\mu=1}^{N}
\int d{\bf k} \,
\bigl\{
(\widehat{\bf q} \cdot {\bf k})^{2}
c_{\lambda a}(k) c_{\mu b}(k)
F_{\lambda \mu}(k,t) F_{ab}(p,t) +
\nonumber \\
& & \qquad \qquad \qquad \quad
+ \,
(\widehat{\bf q} \cdot {\bf k})
(\widehat{\bf q} \cdot {\bf p})
c_{\lambda a}(k) c_{\mu b}(p)
F_{\lambda b}(k,t) F_{a \mu}(p,t)
\bigr\}.
\label{eq:appendix-MCT}
\end{eqnarray}

\subsection{MCT equations for tagged molecule's correlator}
\label{appendix:MCT-3}

The MCT equations of motion for a
tagged molecule (labeled ${\rm s}$)
can be derived in a similar manner, and only the outline
of the derivation and resulting equations shall be
presented in the following.
The Zwanzig-Mori equation for the tagged molecule's
density correlator
$F_{ab}^{\rm s}(q,t) =
\langle \rho_{a}^{\rm s}({\bf q}) e^{i {\cal L} t}
\rho_{b}^{\rm s}({\bf q}) \rangle$
is obtained from the projection operator ${\cal P}^{\rm s}$
onto $\rho_{a}^{\rm s}({\bf q})$
and $j_{a}^{\rm s}({\bf q}) =
v_{{\rm s},z}^{a} e^{i {\bf q} \cdot {\bf r}_{\rm s}^{a}}$,
and is given by
\begin{equation}
\ddot{F}_{ab}^{\rm s}(q,t) +
\sum_{x=1}^{N} \Omega_{ax}^{{\rm s} \, 2}(q) \, F_{xb}^{\rm s}(q,t) +
\sum_{x=1}^{N} \int_{0}^{t} dt' \, M_{ax}^{\rm s}(q,t-t') \,
\dot{F}_{xb}^{\rm s}(q,t') = 0.
\label{eq:appendix-GLE-s}
\end{equation}
Here the characteristic frequency matrix reads
\begin{equation}
\Omega_{ab}^{{\rm s} \, 2}(q) = q^{2} v^{2} w^{-1}_{ab}(q),
\end{equation}
and the formal expression for the memory kernel is given by
\begin{equation}
M_{ab}^{\rm s}(q,t) =
\frac{1}{v^{2}}
\langle f_{a}^{\rm s}({\bf q})^{*}
\exp (i {\cal Q}^{\rm s} {\cal L} {\cal Q}^{\rm s} t)
f_{b}^{\rm s}({\bf q}) \rangle,
\end{equation}
in terms of the fluctuating random force
evolving with ${\cal Q}^{\rm s} {\cal L} {\cal Q}^{\rm s}$
(${\cal Q}^{\rm s} \equiv 1 - {\cal P}^{\rm s}$)
\begin{equation}
f_{a}^{\rm s}({\bf q}) =
\dot{j}_{a}^{\rm s}({\bf q})
- i q v^{2}
\sum_{\nu} \rho_{\nu}^{\rm s}({\bf q}) w_{\nu a}^{-1}(q).
\end{equation}

The memory kernel under the mode-coupling approximation reads
\begin{equation}
M_{ab}^{\rm s}(q,t) =
\frac{1}{v^{2}}
\langle {\cal P}_{2}^{\rm s} f_{a}^{\rm s}({\bf q})^{*}
\exp (i {\cal Q}^{\rm s} {\cal L} {\cal Q}^{\rm s} t)
{\cal P}_{2}^{\rm s} f_{b}^{\rm s}({\bf q}) \rangle.
\label{eq:appendix-MCT-s-P2}
\end{equation}
Here the operator ${\cal P}_{2}^{\rm s}$ 
projects any variable onto the subspace spanned
by pair density modes
$A_{\lambda \mu}^{\rm s}({\bf k}, {\bf p}) =
\rho_{\lambda}^{\rm s}({\bf k}) \rho_{\mu}({\bf p})$
formed by the tagged molecule's density fluctuations
and collective ones.
Adopting the convolution approximation
relevant here~\cite{Chong98}
\begin{equation}
\langle
\rho_{\lambda}^{\rm s}({\bf k})^{*}
\rho_{\mu}({\bf p})^{*}
\rho_{\nu}^{\rm s}({\bf q})
\rangle
\approx
\sum_{\sigma}
w_{\lambda \sigma}(k) \rho h_{\mu \sigma}(p) w_{\nu \sigma}(q),
\end{equation}
one obtains for the projected random force
\begin{eqnarray}
& &
{\cal P}_{2}^{\rm s} f_{\rm s}^{a}({\bf q}) =
-i \frac{\rho v^{2}}{n}
\sum_{\bf k} \sum_{\lambda, \mu, \nu}
\left\{
  k_{z} [ \delta_{\lambda a} - w_{\lambda a}^{-1}(k) ] +
  p_{z} \delta_{\lambda a} 
\right\}
w_{a \nu}(p) c_{\nu \mu}(p)
\rho_{\lambda}^{\rm s}({\bf k})
\rho_{\mu}({\bf p}) 
\nonumber \\
& & \qquad
=
-i \frac{\rho v^{2}}{n}
\sum_{\bf k} \sum_{\lambda, \mu, \nu}
\left\{
  k_{z} \rho c_{\lambda a}^{\rm intra}(k) +
  p_{z} \delta_{\lambda a} 
\right\}
[{\bf I} - \rho {\bf c}^{\rm intra}(p)]^{-1}_{a \nu}
c_{\nu \mu}(p) 
\rho_{\lambda}^{\rm s}({\bf k}) 
\rho_{\mu}({\bf p}) ,
\label{eq:fs-before}
\end{eqnarray}
where in the final equality
the intramolecular structure factor is expressed
in terms of $c^{\rm intra}$. 
Again, intra-molecular contributions 
given in terms of $c^{\rm intra}$ shall be neglected here,
which leads to 
\begin{equation}
{\cal P}_{2}^{\rm s} f_{a}^{\rm s}({\bf q}) =
- \frac{i \rho v^{2}}{n}
\sum_{\bf k}
\sum_{\lambda}
(\widehat{\bf q} \cdot {\bf p}) c_{a \lambda}(p) \,
\rho_{a}^{\rm s}({\bf k})
\rho_{\lambda}({\bf p}).
\label{eq:fs}
\end{equation}
All the theoretical results presented in the main text are
not much affected by this neglect since $w_{ab}(q)$
for microscopic wave vectors, in particular, 
near the peak position $q^{*}$ of the static structure
factor $S(q)$, are close to diagonal~\cite{Aichele04},
in which case there holds $c_{ab}^{\rm intra}(q) \approx 0$.
In addition, our derivation of the Rouse model
presented in Appendix~\ref{appendix:Rouse-Fuchs} is not altered
since the relevant memory kernel there,
$\widehat{m}_{\lambda=0}(t)$, is formed by 
the summation 
$\sum_{a=1}^{N} {\cal P}_{2}^{\rm s} f_{a}^{\rm s}({\vec q} \to 0)$
of the projected random forces in the small wave vector limit,
for which both 
Eqs.~(\ref{eq:fs-before}) and (\ref{eq:fs}) 
yield the identical expression.
Under the factorization approximation
\begin{equation}
\frac{1}{n}
\langle
\rho_{a}^{\rm s}({\bf k})^{*}
\rho_{\lambda}({\bf p})^{*}
e^{i {\cal Q}^{\rm s} {\cal L} {\cal Q}^{\rm s} t}
\rho_{b}^{\rm s}({\bf k})
\rho_{\mu}({\bf p}) \rangle
\approx
F_{ab}^{\rm s}(k,t) \, F_{\lambda \mu}(p,t),
\label{eq:factori-s-appendix}
\end{equation}
substituting Eq.~(\ref{eq:fs}) into 
Eq.~(\ref{eq:appendix-MCT-s-P2}) finally yields
\begin{equation}
M_{ab}^{\rm s}(q,t) =
\frac{\rho v^{2}}{(2 \pi)^{3}}
\sum_{\lambda,\mu=1}^{N}
\int d{\bf k} \,
(\widehat{\bf q} \cdot {\bf p})^{2}
c_{a \lambda}(p) c_{b \mu}(p)
F_{ab}^{\rm s}(k,t) F_{\lambda \mu}(p,t).
\label{eq:appendix-MCT-s}
\end{equation}

\section{Derivation of the Rouse model}
\label{appendix:Rouse-Fuchs}

In this appendix, we show that our microscopic formulation
for the polymer dynamics
based on MCT reduces to the Rouse model in the asymptotic
limit of large degrees of polymerization $N$.
Implications of our theory in this limit
on the Rouse-mode correlators
in the $\beta$ and $\alpha$ relaxation regimes
and possible finite-$N$ corrections are also
discussed.

\subsection{MCT equations for mean-square displacements}
\label{appendix:Rouse-Fuchs-1}

We start from deriving the MCT equations for monomer
MSDs
\begin{equation}
\Delta r_{ab}^{2}(t) \equiv
\delta r_{ab}^{2}(t) - \delta r_{ab}^{2}(0)
\quad \mbox{with} \quad
\delta r_{ab}^{2}(t) =
\langle [ \, {\bf r}_{a}(t) - {\bf r}_{b}(0) \, ]^{2} \rangle.
\end{equation}
Notice that the CM MSD $g_{\rm C}(t)$ and the monomer-averaged
one $g_{\rm M}(t)$ introduced in Sec.~\ref{sec:theory-3}
can be expressed in terms of
$\Delta r_{ab}^{2}(t)$ as
\begin{eqnarray}
g_{\rm C}(t) &=& \frac{1}{N^{2}}
\sum_{a,b=1}^{N} \Delta r_{ab}^{2}(t),
\label{eq:gC-Delta}
\\
g_{\rm M}(t) &=& \frac{1}{N}
\sum_{a=1}^{N} \Delta r_{aa}^{2}(t) =
\frac{1}{N} \, \mbox{Tr}[\, \Delta r^{2}(t) \,].
\label{eq:gM-Delta}
\end{eqnarray}
Since $F_{ab}^{\rm s}(q \to 0,t) = 1 -
q^{2} \delta r_{ab}^{2}(t)/6 + O(q^{4})$
[{\em cf.} Eq.~(\ref{eq:Fs-def})],
the Zwanzig-Mori equation for $\Delta r_{ab}^{2}(t)$ can be derived
from the small-$q$ behavior of Eq.~(\ref{eq:GLE-s})
\begin{equation}
\frac{1}{v^{2}} \partial_{t}^{2} \Delta r_{ab}^{2}(t) +
q^{2} \sum_{x=1}^{N} w_{ax}^{-1}(q \to 0) \, \Delta r_{xb}^{2}(t) +
\sum_{x=1}^{N} \int_{0}^{t} dt' \,
m_{ax}(t-t') \, \partial_{t'} \Delta r_{xb}^{2}(t') = 6 \, \delta_{ab},
\label{eq:GLE-MSD}
\end{equation}
with the memory kernel
$m_{ab}(t) = \lim_{q \to 0} q^{2} \sum_{x}
w_{ax}^{-1}(q) m_{xb}(t)$
[{\em cf.} Eqs.~(\ref{eq:MCT-s}) and (\ref{eq:vertex-s})]:
\begin{equation}
m_{ab}(t) = \frac{\rho_{\rm m}}{6 \pi^{2}}
\int dk \, k^{4} S(k) c(k)^{2} \,
F_{ab}^{\mathrm s}(k,t) \, \phi(k,t).
\label{eq:MCT-MSD}
\end{equation}
In Eq.~(\ref{eq:GLE-MSD}),
the term $q^{2} w_{ab}^{-1}(q \to 0)$ has to be kept
since $w_{ab}(q=0) = 1$ is singular
and its inverse does not exist.
For $g_{\rm C}(t)$,
a somewhat simplified equation can be
derived by taking $(1/N^{2}) \sum_{a,b}$ of Eq.~(\ref{eq:GLE-MSD})
and noticing
$\lim_{q \to 0} q^{2} \sum_{a} w_{ax}^{-1}(q) = 0$:
\begin{equation}
\frac{1}{v^{2}} \partial_{t}^{2} g_{\rm C}(t) +
\frac{1}{N^{2}} \sum_{a,x,b=1}^{N}
\int_{0}^{t} dt' \,
m_{ax}(t-t') \, \partial_{t'} \Delta r_{xb}^{2}(t') = \frac{6}{N} \, .
\label{eq:GLE-gC}
\end{equation}

\subsection{Derivation of the Rouse model as asymptotic solution}
\label{appendix:Rouse-Fuchs-2}

In general no simplification of the complicated couplings in the
motions of all monomers is possible, as an exact diagonalization of
the integro-differential equations for the matrix $\Delta
r^{2}_{ab}(t)$ is required. Only for long times and large degrees of
polymerization $N$, an asymptotic solution is possible and is
presented in the following. It rests upon the property of the memory
functions $m_{ab}(t)$ in Eq.~(\ref{eq:MCT-MSD}) that they are
`cut-off' by the collective density fluctuations. The slowest
collective correlator $\phi(q,t)$ is connected with the average
monomer separation and lies at the position $q^{*}$ of the peak of
$S(q)$ ({\em cf.} Sec.~\ref{sec:results-2}). Thus, at long times,
the tagged polymer's density correlator $F^{\rm s}_{ab}(q,t)$ at the
(asymptotically $N$-independent) wave vector $q^{*}$ dominates the
memory functions. It is bounded by the intrachain structure factor
at that wave vector, i.e., $w_{ab}(q^{*}) > F_{ab}^{\rm s}(q^{*},t)$
[in the sense that $w_{ab}(q^{*}) - F_{ab}^{\rm s}(q^{*},t)$ is
positive definite]. As long as $w_{ab}(q^{*})$ on this length scale
contains no anomalous correlations extending over large monomer
separations, i.e., $w_{ab}(q^{*}) \to 0$ for $| \, a-b \, | \to
\infty$, the same property holds for the memory function as well:
$m_{ab}(t)\to 0$ for $| \, a-b \, | \to \infty$. This property, and
that $m_{ab}(t)$ decays to zero for times longer than
$\tau_{q^{*}}$, the structural relaxation time, are the central
ingredients to the derivation of the Rouse model within our
approach. 
Note that the above reasoning also holds if the role
of the microscopic wave vector $q^{*}$ is replaced by some other
wave vector $q^{\dagger}$, as long as $q^{\dagger}$ is characteristic of
local motion and asymptotically $N$-independent. 
Thus, the following
reasoning also applies to high temperature, where, as discussed in
Sec.~\ref{sec:results-3-3},
the slowest collective mode in the simulation result is found
to lie at $q \approx q_{\rm C}$~\cite{comment-qC-for-various-N}.

For long times, $t\to\infty$,
Eq.~(\ref{eq:GLE-gC}) is solved by a uniform increase
of all MSDs following the CM motion:
\begin{equation}
g_{\rm C}(t) \to 6 D t
\quad \mbox{and} \quad
\delta r^{2}_{ab}(t) \to g_{\rm C}(t) + O(t^{x}),
\end{equation}
where the Markovian limit in the memory function
requires $t\gg \tau_{q^{*}}$.
Below, we will determine the leading correction
that exhibits a power-law behavior with the
exponent $x=1/2$.
As explained above, the site dependence of $m_{ab}(t)$
for long times is dominated by the tagged polymer's
density fluctuations at microscopic wave vectors.
Therefore, the summation over site indices and
consecutive $k$-integration will asymptotically become
$N$-independent, and the diffusion constant $D$ scales like
\begin{equation}
D = \frac{k_{\rm B}T}{N \zeta} \quad (N \to \infty),
\label{eq:diffusion}
\end{equation}
with the (asymptotically $N$-independent)
friction coefficient $\zeta$ determined by
\begin{equation}
\zeta / k_{\rm B}T =
\frac{\rho_{\rm m}}{6\pi^{2}} \int_{0}^{\infty}dt \int dk \,
k^{4} S(k) c^{2}(k)
\biggl[ \, \frac{1}{N} \sum_{a,b=1}^{N} F^{\rm s}_{ab}(k,t) \, \biggr] \,
\phi(k,t).
\label{eq:zeta}
\end{equation}
We confirmed 
that the Gaussian chains studied in Ref.~\cite{Chong02c}
follows the asymptotic law (\ref{eq:diffusion}) for large $N$.

In considering internal-mode contributions to the monomer MSD in the
limit of $N \to \infty$, chain-end effects can be neglected, and the
structure of the various matrices in Eq.~(\ref{eq:GLE-MSD})
simplifies. We can assume that they depend only on the difference of
indices, $s=a-b$ (characteristic of `Toeplitz'
matrices~\cite{Grenander58}), and we define, e.g., $\Delta
r_{(s=a-b)}^{2}(t) = \Delta r_{ab}^{2}(t)$. This assumption neglects
monomer correlations caused by chain ends and does not hold, e.g.,
for isolated self-avoiding-walk polymers in good solvents whose end
regions are slightly less swollen than middle portions. 
Similar correlations have recently been discovered
also in long chain polymer melts, but the amplitude
is much weaker than in dilute solution~\cite{Wittmer07b}.
Here, as a first step, we
assume that such non-trivial correlations to be absent.

For the monomer-averaged MSD, it suffices to obtain the
distribution of the internal modes since $g_{\rm M}(t)$
in Eq.~(\ref{eq:gM-Delta}) is expressed as the trace of
the matrix $\Delta {r}_{ab}^{2}(t)$,
where only the eigenvalues of the internal modes enter.
The distribution in the $N \to \infty$
limit can be found by assuming periodic
boundary conditions and performing a Fourier transform.
Transformed quantities like
\begin{equation}
\widehat{\Delta r}^{2}_{\lambda}(t) =
\sum_{s=-\infty}^{\infty}  e^{i \lambda s} \Delta r_{(s)}^{2}(t),
\end{equation}
shall be marked by a hat. Notice that there hold $g_{\rm C}(t) =
(1/N) \widehat{\Delta r}^{2}_{\lambda=0}(t)$ and $g_{\rm M}(t) = (1/N)
\sum_{\lambda} \widehat{\Delta r}^{2}_{\lambda}(t)$, so that the
internal-mode contribution to the monomer-averaged MSD is given by
$g_{\rm M}(t) - g_{\rm C}(t) = (1/N) \sum_{\lambda \ne 0}
\widehat{\Delta r}^{2}_{\lambda}(t)$. In the asymptotic $N \to \infty$
limit, the monomer-averaged MSD follows from the density of states
of internal modes via~\cite{Grenander58}
\begin{equation}
g_{\rm M}(t) - g_{\rm C}(t) \to
\int_{-\pi}^{\pi} \frac{d\lambda}{2\pi} \,
\widehat{\Delta r}^{2}_{\lambda}(t).
\label{eq:gMC}
\end{equation}
Here and in the following, $\lambda \ne 0$ shall be
assumed unless stated otherwise.

The equation of motion for $\widehat{\Delta r}^{2}_{\lambda}(t)$ is
obtained from Eq.~(\ref{eq:GLE-MSD}) via Fourier transformation,
recognizing that matrix products, owing to the assumption of the
dependence on the index-difference only, become convolution and turn
into simple products after Fourier transformation:
\begin{equation}
\frac{1}{v^{2}}
\partial_{t}^{2} \widehat{\Delta r}^{2}_{\lambda}(t) +
\widehat{\Gamma}_{\lambda} \widehat{\Delta r}^{2}_{\lambda}(t) +
\int_{0}^{t} dt' \,
\widehat{m}_{\lambda}(t-t') \,
\partial_{t'} \widehat{\Delta r}^{2}_{\lambda}(t') = 6.
\label{eq:GLE-Delta-MSD-lambda}
\end{equation}
Here we have introduced
\begin{equation}
\widehat{\Gamma}_{\lambda}^{-1} \equiv
\lim_{q \to 0} \frac{\widehat{w}_{\lambda}(q)}{q^{2}} =
-\frac{1}{6} \sum_{s=-\infty}^{\infty}
e^{i \lambda s} \delta r^{2}_{(s)}(0),
\label{eq:Gamma-def}
\end{equation}
and the transformed memory kernel is given by
\begin{equation}
\widehat{m}_{\lambda}(t) = \frac{\rho_{\rm m}}{6\pi^{2}}
\int dk \, k^{4} S(k) c^{2}(k) \widehat{F}^{\rm s}_{\lambda}(k,t)
\phi(k,t).
\label{eq:MCT-Delta-MSD-lambda}
\end{equation}

Equations (\ref{eq:gMC}) and (\ref{eq:GLE-Delta-MSD-lambda}) yield
the $N$-independent growth of the monomer-averaged MSD resulting
from the internal modes, whose spectrum shall be determined
to lowest order in the mode parameter $\lambda$.
The Gaussian approximation shall be assumed for the large
separation behavior in $\delta r^{2}_{(s=a-b)}(0)$~\cite{Wittmer07b}, 
so that the small $\lambda$ properties of $\widehat{\Gamma}_{\lambda}$
can be found from
\begin{equation}
\widehat{\Gamma}_{\lambda}^{-1} \approx -\frac{1}{6}
\sum_{s=-\infty}^{\infty} e^{i \lambda s} \, |\, s \, | \,
\sigma_{\rm s}^{2}
\to \frac{\sigma_{\rm s}^{2}}{3 \lambda^{2}} \, .
\label{eq:Gamma-expansion}
\end{equation}
Here $\sigma_{\rm s}$ denotes the statistical segment length.
At the same time, the memory kernel is Taylor-expanded
for small $\lambda$, and then a Markovian approximation
({\em cf.} the next subsection)
is performed
\begin{equation}
\widehat{m}_{\lambda}(t) \approx
\biggl[ \,
  \int_{0}^{\infty} dt \, \widehat{m}_{\lambda=0}(t) \,
\biggr] \, \delta(t)
= (\zeta/k_{\rm B}T) \, \delta(t),
\label{eq:memory-Markov}
\end{equation}
to derive the long time behavior.
It is justified, as discussed above, because the memory kernel
is dominated by microscopic wave vectors, where
the summation over $s=a-b$ converges rapidly and an
expansion in $\lambda$ is possible.
In Eq.~(\ref{eq:memory-Markov}) only the lowest order in $\lambda$
is retained, which is given by the friction coefficient
from Eq.~(\ref{eq:zeta}) of the CM motion.
Neglecting the inertia term for long times, we find
from Eq.~(\ref{eq:GLE-Delta-MSD-lambda})
for the small-$\lambda$ modes
\begin{equation}
\frac{\zeta}{k_{\rm B}T} \, \partial_{t} \widehat{\Delta r}^2_{\lambda}(t) +
\frac{3 \lambda^{2}}{\sigma_{\rm s}^{2}} \, \widehat{\Delta r}_{\lambda}^{2}(t) = 6,
\end{equation}
with the initial value $\widehat{\Delta r}^{2}_{\lambda}(0) = 0$,
or equivalently
\begin{equation}
\frac{\zeta}{k_{\rm B}T} \, \partial_{t} 
\widehat{\delta r}^{2}_{\lambda}(t) + 
\frac{3 \lambda^{2}}{\sigma_{\rm s}^{2}} \,
\widehat{\delta r}^{2}_{\lambda}(t) = 0, 
\label{eq:anomaly-a}
\end{equation}
whose solution reads
\begin{equation}
\widehat{\delta r}^{2}_{\lambda}(t) =
e^{-3 k_{\rm B}T \lambda^{2} t / \zeta \sigma_{\rm s}^{2}} \,
\widehat{\delta r}^{2}_{\lambda}(0),
\label{eq:anomaly-b}
\end{equation}
with $\widehat{\delta r}^{2}_{\lambda}(0)= - 2 \sigma_{\rm s}^{2}/\lambda^2$.
From this, the following monomer-averaged MSD follows for long times
as familiar in the Rouse model~\cite{Doi86}:
\begin{equation}
g_{\rm M}(t) - g_{\rm C}(t) = \int_{-\pi}^{\pi}
\frac{d\lambda}{2\pi} \, \frac{\sigma_{\rm s}^{2}}{2\lambda^{2}}
\left( 1 - e^{-3 k_{\rm B}T \lambda^{2} t / \zeta \sigma_{\rm
s}^{2}} \right) \to \frac{2 \sigma_{\rm s}^{2}}{\pi^{3/2}} \, \sqrt{
\frac{3 \pi^{2} k_{\rm B}T}{\zeta \sigma_{\rm s}^{2}} } \, \sqrt{t}.
\label{eq:anomaly}
\end{equation}
This concludes the derivation of the Rouse model as the asymptotic
large chain-length limit of the MCT equations for a polymer chain 
dissolved in a melt of identical polymers. 
(The use of the Markovian approximation will be justified
in the next subsection.)
We find the expected scaling
of the diffusion coefficient with molecular weight in
Eq.~(\ref{eq:diffusion}), the (low-lying) spectrum of eigenvalues in
Eq.~(\ref{eq:anomaly-b}), and the resulting anomaly in the monomer
MSD, Eq.~(\ref{eq:anomaly}). The occurring parameters can be measured
from global chain properties: the friction
coefficient $\zeta$ from the averaged friction kernel,
Eq.~(\ref{eq:zeta}), and the segment length $\sigma_{\rm s}$ from the
Gaussian behavior of the equilibrium segment correlations at large
separation, Eq.~(\ref{eq:Gamma-expansion}).

\subsection{Implications on the Rouse-mode correlators}
\label{appendix:Rouse-Fuchs-3}

We notice that $\widehat{\delta r}^{2}_{\lambda}(t)$
is essentially the (diagonal) Rouse-mode correlators
$C_{pp}(t)$ introduced in Sec.~\ref{sec:theory-3}
with the correspondence $\lambda \sim p/N$
since there holds
$6 \sum_{p \ge 1} [ C_{pp}(0) - C_{pp}(t) ] =
\sum_{\lambda \ne 0}
[ \widehat{\delta r}^{2}_{\lambda}(t) - \widehat{\delta r}^{2}_{\lambda}(0) ]$
from the comparison of 
Eqs.~(\ref{eq:MSD-monomer-averaged}) and (\ref{eq:gMC}).
The difference comes from the boundary condition adopted in
defining $C_{pp}(t)$ and $\widehat{\delta r}^{2}_{\lambda}(t)$.
So, the small-$p$ properties of the normalized
Rouse-mode correlators
$c_{p}(t) = C_{pp}(t)/C_{pp}(0)$ can be deduced
from the small-$\lambda$ behavior of
\begin{equation}
\widehat{c}_{\lambda}(t) =
\widehat{\delta r}^{2}_{\lambda}(t) \, / \, 
\widehat{\delta r}^{2}_{\lambda}(0).
\label{eq:c-lambda-def}
\end{equation}
The equation of motion for $\widehat{c}_{\lambda}(t)$
can be derived from Eq.~(\ref{eq:GLE-Delta-MSD-lambda}):
\begin{equation}
\frac{1}{v^{2}}
\partial_{t}^{2} \widehat{c}_{\lambda}(t) +
\widehat{\Gamma}_{\lambda} \widehat{c}_{\lambda}(t) +
\int_{0}^{t} dt' \,
\widehat{m}_{\lambda}(t-t') \,
\partial_{t'} \widehat{c}_{\lambda}(t') = 0.
\label{eq:GLE-c-lambda}
\end{equation}
The Laplace transform of this equation reads
\begin{equation}
- \frac{1}{v^{2}}
z [ 1 + z \widehat{c}_{\lambda}(z) ] +
\widehat{\Gamma}_{\lambda} \widehat{c}_{\lambda}(z) -
\widehat{m}_{\lambda}(z) \,
[ 1 + z \widehat{c}_{\lambda}(z) ] = 0,
\label{eq:Laplace-GLE-c-lambda}
\end{equation}
where the convention
$f(z) = i \int_{0}^{\infty} dt \, e^{izt} f(t)$
with ${\rm Im} \, z > 0$ is adopted.

Let us consider liquid states, for which there is no
nonergodicity pole in 
the Laplace transform of correlators.
(Nonergodicity parameters shall be discussed below.)
Then, we have from the $z \to 0$ limit of
Eq.~(\ref{eq:Laplace-GLE-c-lambda}):
$\widehat{\Gamma}_{\lambda} \widehat{c}_{\lambda}(z \to 0) =
\widehat{m}_{\lambda}(z \to 0)$.
Retaining only the leading-order contribution in
$\lambda$ for $\widehat{\Gamma}_{\lambda}$ and
$\widehat{m}_{\lambda}$, one obtains for small $\lambda$
\begin{equation}
\widehat{c}_{\lambda}(z \to 0) =
\frac{\sigma_{\rm s}^{2}}{3 \lambda^{2}} \,
\widehat{m}_{\lambda=0}(z \to 0).
\label{eq:time-scale-separation}
\end{equation}
Since $\widehat{c}_{\lambda}(z \to 0)$ and
$\widehat{m}_{\lambda=0}(z \to 0)$ are proportional to
their relaxation times, this implies that
the relaxation time of $\widehat{c}(t)$
[and hence of $\widehat{\delta r}^2_{\lambda}(t)$]
is larger by a factor of
$1/\lambda^{2} \sim (N/p)^{2}$
than that of $\widehat{m}_{\lambda=0}(t)$.
This justifies the use of the Markovian approximation,
which has been adopted in Eq.~(\ref{eq:memory-Markov}).
Since, as discussed above, the relaxation time of
$\widehat{m}_{\lambda=0}(t)$ is dictated by 
that of the 
coherent dynamics at microscopic wave vectors,
Eq.~(\ref{eq:time-scale-separation})
also implies the separation of the time scale
for $\widehat{c}(t)$ from that for the
density fluctuations at the monomer length scale.

We next turn our attention to nonergodicity parameters
$\widehat{f}_{\lambda} = \widehat{c}_{\lambda}(t \to \infty)$
and
$\widehat{m}_{\lambda} = \widehat{m}_{\lambda}(t \to \infty)$
in glass states.
Since there holds
$\lim_{t \to \infty} f(t) = - \lim_{z \to 0} z f(z)$,
one obtains from the $z \to 0$ limit of
Eq.~(\ref{eq:Laplace-GLE-c-lambda}):
$\widehat{f}_{\lambda} = \widehat{m}_{\lambda} /
[\widehat{m}_{\lambda} + \widehat{\Gamma}_{\lambda}]$.
Because of Eq.~(\ref{eq:Gamma-expansion}),
we have
$\widehat{f}_{\lambda} = 1$ in leading order for small $\lambda$.
This holds also at the MCT critical point, so that
the critical nonergodicity parameter for
$\widehat{c}_{\lambda}(t)$ is $\widehat{f}_{\lambda}^{\rm c} = 1$.
Since $\widehat{c}_{\lambda}(t) \le 1$, this also implies
that the critical amplitude for $\widehat{c}_{\lambda}(t)$ is zero,
$\widehat{h}_{\lambda} = 0$.
Therefore, our theory predicts that,
for small mode indices or in the $N \to \infty$ limit,
the critical nonergodicity parameter of the Rouse-mode
correlators is unity, and they do not exhibit
the MCT $\beta$ dynamics ({\em cf.} Sec.~\ref{sec:theory-4}).

Let us then consider the dynamics of $\widehat{c}_{\lambda}(t)$
in the $\alpha$ regime, which can be described by its
$\alpha$ master curve
[{\em cf.} Eq.~(\ref{eq:superposition})].
We consider the dynamics on the $\alpha$-relaxation time
scale $t_{\sigma}^{\prime}$ [{\em cf.} Eq.~(\ref{eq:t-sigma-prime})]
and write, e.g.,
$\widehat{c}_{\lambda}(t) = \tilde{c}_{\lambda}(\tilde{t})$ with
$\tilde{t} = t / t_{\sigma}^{\prime}$ and
$\widehat{c}_{\lambda}(z) = t_{\sigma}^{\prime}
\tilde{c}_{\lambda}(\tilde{z})$ with
$\tilde{z} = z t_{\sigma}^{\prime}$,
with a yet unspecified function $\tilde{c}_{\lambda}$.
It then follows from Eq.~(\ref{eq:Laplace-GLE-c-lambda}):
\begin{equation}
- \frac{1}{v^{2}} \,
\frac{\tilde{z}}{(t_{\sigma}^{\prime})^{2}} \,
[ 1 + \tilde{z} \tilde{c}_{\lambda}(\tilde{z}) ] +
\widehat{\Gamma}_{\lambda} \tilde{c}_{\lambda}(\tilde{z}) -
\tilde{m}_{\lambda}(\tilde{z}) \,
[ 1 + \tilde{z} \tilde{c}_{\lambda}(\tilde{z}) ] = 0.
\end{equation}
Now, the $\alpha$-scaling limit shall be
performed:
$t_{\sigma}^{\prime} \to \infty$ for
$T \to T_{\rm c}+$, but
with $\tilde{t}$ and $\tilde{z}$ fixed~\cite{Goetze91b}.
We thus obtain
$\widehat{\Gamma}_{\lambda} \tilde{c}_{\lambda}(\tilde{z}) -
\tilde{m}_{\lambda}(\tilde{z}) \,
[ 1 + \tilde{z} \tilde{c}_{\lambda}(\tilde{z}) ] = 0$.
Since $\tilde{c}_{\lambda}(\tilde{t} \to 0) =
\widehat{f}_{\lambda}^{\rm c}$~\cite{Goetze91b}
and $\widehat{f}_{\lambda}^{\rm c} = 1$ for small $\lambda$
as derived above, the inverse Laplace transform of this
equation yields
\begin{equation}
\widehat{\Gamma}_{\lambda} \tilde{c}_{\lambda}(\tilde{t}) +
\int_{0}^{\tilde{t}} d\tilde{t}' \,
\tilde{m}_{\lambda}(\tilde{t}-\tilde{t}')
\partial_{\tilde{t}'}
\tilde{c}_{\lambda}(\tilde{t}') = 0.
\label{eq:alpha-master-tilde-c}
\end{equation}
Again, only the leading-order contribution in $\lambda$
for $\widehat{\Gamma}_{\lambda}$ and $\tilde{m}_{\lambda}$
shall be retained.
With the same reasoning as presented concerning
Eq.~(\ref{eq:time-scale-separation}),
one obtains the time-scale separation of the dynamics
of $\tilde{c}_{\lambda}(\tilde{t})$ from that of
$\tilde{m}_{\lambda=0}(\tilde{t})$.
Therefore, the Markovian approximation for
Eq.~(\ref{eq:alpha-master-tilde-c}) is justified,
leading for small $\lambda$ to
the exponential decay of the $\alpha$ master curve
\begin{equation}
\tilde{c}_{\lambda}(\tilde{t}) = \exp[ - \tilde{t} /
\tilde{\tau}_{\lambda}],
\end{equation}
with the relaxation time
$\tilde{\tau}_{\lambda} =
\tilde{\zeta} \sigma_{\rm s}^{2} \, / \, 3 k_{\rm B} T \lambda^{2}$
($\tilde{\zeta} = \lim_{T \to T_{\rm c}+}
\zeta / t_{\sigma}^{\prime}$)
whose dependence on the mode index reads
$1/\lambda^{2} \sim (N/p)^{2}$.
Thus, our theory in the $N \to \infty$ limit
yields the Rouse-model result for the $\alpha$-master curves
for the Rouse-mode correlators.

\subsection{Finite-N corrections}
\label{appendix:Rouse-Fuchs-4}

So far, we have derived the asymptotic solution of our MCT equations
in the limit of large degrees of polymerization $N$,
by retaining only the leading contribution in the
expansion in the mode parameter.
To find finite-$N$ corrections, one has to go
beyond the leading order,
but it is difficult to explicitly work this out.
On the other hand,
it is obvious that finite-$N$ corrections
lead to deviations from the asymptotic (Rouse model) results:
when the normalized Rouse-mode correlator $c_{p}(t)$
in the $\alpha$ regime is fitted via a
Kohlrausch function
$A_{p} \exp[ - (t / \tau_{p})^{\beta_{p}}]$,
finite-$N$ corrections lead to
$A_{p} < 1$, $\beta_{p} < 1$, and
deviations from
$\tau_{p} \propto (N/p)^{2}$.
Furthermore, the monomer-averaged MSD does not exhibit the
square-root-time dependence any longer.

\newpage

\begin{thebibliography}{44}
\bibitem[{com({\natexlab{b}})}]{comment-equivalent-site-app}
\bibinfo{note}{We have explicitly compared the dynamics resulting with and
  without the use of the equivalent-site approximation (\ref{eq:PRISM-c}) for
  short Gaussian chains where chain-end effects are more prominent. No
  substantial differences have been found.}
\bibitem[{com({\natexlab{c}})}]{comment-computational-time}
\bibinfo{note}{Without any further approximation introduced in the main text, a
  typical CPU time requested to solve the full matrix MCT equations
  (\ref{eq:GLE}) and (\ref{eq:MCT}) for the present model ($N=10$) at a
  temperature near the MCT critical point $T_{\rm c}$ is estimated to be of the
  order of one month on a Pentium IV processor with 2.80 GHz clock speed.}
\bibitem[{com({\natexlab{d}})}]{comment-regular-part}
\bibinfo{note}{The regular contribution, which is not taken into account in our
  theory, might contain some polymer-specific (e.g., chain-length dependent)
  features.}
\bibitem[{com({\natexlab{e}})}]{comment-coherent-vs-incoherent}
\bibinfo{note}{For example, the long-time limit of the coherent density
  correlator $f(q) \equiv \phi(q,t \to \infty)$ can be determined by solving
  Eq.~(\ref{eq:NEP}), and one has to solve analogous matrix equations for
  obtaining that of the single-chain density correlators $F_{ab}^{s}(q,t \to
  \infty)$. The solution of Eq.~(\ref{eq:NEP}) can be found numerically by an
  iterative procedure $f^{(j+1)}(q)/(1-f^{(j+1)}(q)) = {\cal F}_{q}[f^{(j)}]$
  starting with $f^{(0)}(q) = 1$, and is obtained as $f(q) = \lim_{j \to
  \infty} f^{(j)}(q)$. The convergence of this iterative procedure is ruled by
  the spectral radius of a so-called stability matrix which can be defined in
  terms of the mode-coupling functional ${\cal
  F}_{q}$~\protect\cite{Goetze91b}. The stability matrix attains the maximum
  eigenvalue unity at the MCT critical point $T_{\rm c}$, and more iterations
  are necessary to solve Eq.~(\ref{eq:NEP}) closer to $T_{\rm c}$. However,
  this does not apply to the corresponding iterative procedure for solving the
  equations for the single-chain density correlators, and much less number of
  iterations is sufficient for obtaining $F_{ab}^{s}(q,t \to \infty)$.}
\bibitem[{com({\natexlab{f}})}]{comment-qC-for-various-N}
\bibinfo{note}{We confirmed from our independent computer simulations performed
  for a bead-spring model similar to the one studied here that the wave vector
  at which the high-temperature collective dynamics is slowest does not shift
  with $N$.}
\end{thebibliography}
\end{document}